\RequirePackage{fix-cm}
\documentclass[twocolumn,epjc3]{svjour3}  
\smartqed  
\RequirePackage{color,graphicx,url}
\usepackage{amsfonts,amsmath,amssymb,bm,xspace}
\usepackage{lipsum}
\usepackage[T1]{fontenc} 
\usepackage{epsfig}  
\usepackage{graphicx}  
\usepackage{enumerate}
\usepackage{amssymb}
\usepackage{amsmath}
\usepackage{pifont}
\usepackage[english]{babel}
\usepackage{color}  
\usepackage{multirow}
\definecolor{darkgreen}{rgb}{0.2,0.5, 0.2}
\usepackage{lipsum}
\usepackage{mathtools}
\usepackage{cuted}
\usepackage{dcolumn,booktabs}
\usepackage{ulem}
\newcolumntype{d}[1]{D{.}{.}{#1}}
\usepackage{soul}
\usepackage{xcolor}
\usepackage{graphicx} 
\usepackage{amssymb} 
\usepackage{amsmath}
\usepackage{bbm}
\usepackage{braket} 
\usepackage{makecell}
\usepackage{appendix}
\usepackage{algpseudocode}
\usepackage{xcolor}
\usepackage{color}
\RequirePackage[numbers,sort&compress]{natbib}
\RequirePackage[breaklinks,colorlinks,citecolor=blue,urlcolor=blue,linkcolor=blue]{hyperref}

\def\ga{\,\,\raise0.14em\hbox{$>$}\kern-0.76em\lower0.28em\hbox  
{$\sim$}\,\,}  
\def\la{\,\,\raise0.14em\hbox{$<$}\kern-0.76em\lower0.28em\hbox  
{$\sim$}\,\,}  
\journalname{Eur. Phys. J. A}

\begin{document}

\title{Two-center harmonic oscillator basis for Skyrme-DFT calculations (I): formalism and Proof of Principle}


\author{
        Adri\'{a}n S\'{a}nchez-Fern\'{a}ndez\thanksref{e1,addr1,addr3,addr4} 
        \and
        Jacek Dobaczewski\thanksref{addr1,addr2} 
        \and
        Xuwei Sun\thanksref{addr1} 
        \and
        Herlik Wibowo\thanksref{addr1} 
}
\thankstext{e1}{e-mail: adrian.sanchez.fernandez@ulb.be}

\institute{School of  Physics, Engineering and Technology, University of York,  Heslington, York YO10 5DD, United Kingdom \label{addr1}  \and
Institut d’Astronomie et d’Astrophysique, Université Libre de Bruxelles, Brussels, Belgium \label{addr3} \and
Brussels Laboratory of the Universe - BLU-ULB \label{addr4} \and
Institute of Theoretical Physics, Faculty of Physics, University of Warsaw, ul. Pasteura 5, PL-02-093 Warsaw, Poland \label{addr2}
}

\date{Received: date / Accepted: date}
\maketitle
\begin{abstract}
We present a new method to solve the nuclear density functional theory (DFT) equations using a two-center harmonic oscillator for Skyrme-like functionals, incorporating pairing and Coulomb interactions. The goal is to efficiently determine the fission and fusion configurations in nuclei. The Coulomb exchange term is evaluated exactly, allowing for a novel approach to neck formation without the Slater approximation, commonly used in space coordinate-based approaches. The new method has been implemented in the code {\sc hfodd}, enabling direct comparison with standard one-center solutions. This first paper focuses on deriving and implementing a methodology based on stable, precise, and exact applications of harmonic oscillator bases for the two fragments, which can either overlap or be separated by arbitrarily large distances. The implementation is tested on two proof-of-principle examples using light nuclei, specifically, $^8$Be and $^{24}$Mg.
\end{abstract}
%

%

\section{Introduction}
\label{intro}

After over eighty years, fission is still a fascinating and hot research topic. Heavy atomic nuclei belong to the class of mesoscopic systems that exhibit emergent phenomena, which are difficult to describe using fundamental interactions between nucleons directly. Moreover, contrary to various nuclear properties, which a smaller set of valence nucleons can explain, all nucleons are simultaneously involved in fission. Therefore, fission phenomena are often described in terms of phenomenological models neglecting nucleonic degrees of freedom, hybrid microscopic-macroscopic models, or fully microscopic models based on nuclear density functional theory (DFT)~\cite{(And18),(Sch18b),Krappe2012,Schunck2016,Bender2020,Schunck2022}.

From a microscopic perspective, nuclear fission provides an ideal laboratory for testing and refining our understanding of quantum many-body systems and phenomena. Advancing our theoretical grasp of fission deepens our knowledge of nuclear reactions like fusion and $\alpha-$decay and has broader implications, benefiting fields such as stellar evolution, energy production, or quantum entanglement.

It was recently shown that angular momenta carried by fission fragments are intrinsically related to mass and charge distributions after scission. Even though the experimental data and statistical models seem to corroborate that idea~\cite{Wilson2021}, recent results showed a strong dependency on the scission configurations~\cite{Scamps2022}. Moreover, we can find various articles that provide explanatory models considering even earlier stages in the fission path. For instance, a time-dependent calculation with triple angular momentum projection (light and heavy fragments and the relative motion) shows that the pre-scission bending mode can predict the angular momentum of the fission fragments in agreement with the experimental results~\cite{(Bul22)}. Alternatively, it was also demonstrated that incorporating shell and deformation effects in the moments of inertia of the fragments leads to similar conclusions~\cite{(Ran21)}. Hence, even though statistical models give a valuable bulk approach to the problem, these may fail where more sophisticated microscopic phenomena are at play. This limitation was recently highlighted in the computational model FREYA~\cite{FREYA}, which does not consider the proper $K$-distributions~\cite{(Sca24),(Zho23a)}, one of the key ingredients needed to obtain the relative spin angle between fragments, directly connected to the experiment.

Apart from the angular momentum distribution, it is well-known that many of the fragment properties --such as mass distributions or excitation energies-- are rather sensible to the scission configuration~\cite{Schunck2016}. However, in the existing DFT approaches to fission, relying on the adiabatic approximation, the definition of the scission point is more arbitrary than physical. We need to describe an excited nucleus that is strongly deformed and generally triaxial, and that can be oriented differently in space. Most importantly, we need to link that system to the nascent fragments. In other words, it would be desirable to have a framework that considers the initial nucleus internally structured in fragments, even before the journey through the fission path. In this sense, combining the well-tested DFT models with a two-center basis for building the single-particle (s.p.) states appears to be a meaningful choice.

In recent decades, we have seen two approaches employing the two-center method. The first one relies on extending an external potential (usually Wood-Saxon) into a two-center version~\cite{Sch1971,Chandra78} and solving it on a two-center basis; the harmonic oscillator (HO) is the usual choice. After this, the Schrödinger equation is solved using the Green's function method ~\cite{(Dia08)}. Even though the results shown for both light ~\cite{(Hag17),Oer1996} and heavy systems ~\cite{Maruhn1972,(Mir08)} look promising, the Coulomb interaction is not accurately treated. Indeed, in some cases, it is approximated as a charged spherical distribution, which does not treat overlapping fragments properly. The computational burden associated with obtaining Green's function for large basis sets and the bad solvability conditions when s.p.\ crossings occur make this approach optimal for light masses but cumbersome when describing the evolution of heavy nuclei along the fission path. 

\sloppy
The second type of approach has a stronger ``molecular flavor'' as the main idea is expanding the s.p.\ wave function into two different centers (usually the HO states), in the same way as molecular orbitals are expanded into atomic orbitals, see, e.g., \ Ref.~\cite{(Oer06)}. In nuclear structure studies, results exist for the $\alpha$ clustering, based on hybridization and covalent binding ~\cite{(Fre18)}. In nuclear reactions, adding molecular continuum states improves the description of the scattering of weakly-bound nuclei ~\cite{(Mos21)}. Although many applications exist, these methods are primarily suitable for light systems.

\sloppy
Density Functional Theory (DFT) stands out in pursuing suitable approaches for heavy nuclei, although only a few groups utilized the two-center HO (TCHO) basis expansion. Since the pioneering work of Berger and Gogny more than 40 years ago~\cite{Berger1980}, see also~\cite{(Ber85)}, a handful of axial applications for the Gogny functional exist in areas such as mass distributions~\cite{Dubray2008,Reigner2016,Reigner2019}, fission barriers, and spontaneous fission lifetimes~\cite{Berger1989,Lemaitre2018}. However, the Gogny implementation has never been published, leaving its applicability largely unknown. In addition, the Coulomb interaction was approximated by point-like charges near the scission point~\cite{Berger1989}, which may be inadequate for accurately describing the angular momentum generation of fission fragments. In Covariant DFT, more details are available on the TCHO basis method~\cite{LiSheng2007,Li2023}, although this approach relies on axially deformed co-axial basis states. So far, no implementation of the general framework of the triaxial, shifted, and non-co-axial TCHO basis expansions exist. 

While this strategy has been explored in molecular physics~\cite{Maintz2016}, and some analytical expressions can be applied to the two-center expansion~\cite{Nazmi1996}, no nuclear DFT code exploits this formalism. 

This work focuses on the fundamentals of the two-center (Cartesian) HO basis method and its implementation, which is suitable for the Skyrme energy density functional. We provide detailed information on how to compute the TCHO matrix elements relying on numerical integration instead of transformation of coefficients~\cite{Robledo2022} and how to address the generalized eigenvalue problem in the Hartree-Fock (HF) and HF-Bogoliubov (HFB) cases. This method has been integrated into the latest version of the DFT solver {\sc hfodd}~\cite{(Dob21f),(Dob24a)}, where users can specify the deformation and separation of the bases as input data to perform the HF or HFB calculations.

The structure of the paper is as follows. In section \ref{section2}, we present the theoretical framework, highlighting how the usual Skyrme+Coulomb one-center HO (OCHO) self-consistent procedure can be reformulated to be used within the TCHO basis. In particular, we show how to treat the Coulomb interaction exactly, both in direct and exchange channels. Appendices~A--C present all details of the implementation. In sections \ref{section3} and \ref{section4}, we discuss the results of simple calculations performed for small systems, namely, for $^8$Be, where the Coulomb interaction is tested, and for the symmetric fission channel of $^{24}$Mg. Both Proof-of-Princple TCHO calculations are compared against the standard OCHO results. In section \ref{section5}, we provide final remarks and discuss potential applications of this newly developed method in future research.

\section{Theoretical framework}\label{section2}

In this section, we introduce the TCHO basis states and describe the procedure for solving the Hartree-Fock equations using that basis. Due to its relevance to fission, we also present the method for determining the Coulomb energy in the direct and exchange channels.
 All new features were implemented in the code {\sc hfodd}~\cite{(Dob21f),(Dob24a)} and we refer the reader to the first publication of the code~\cite{Dob1997}, where the corresponding implementation of the OCHO Cartesian basis was defined.

\subsection{The two-center harmonic oscillator basis}
\label{section2a}

We first consider the co-axial case of the TCHO,
 where the principal axes of the two bases coincide, and the bases are shifted by a vector coinciding with one of those principal axes. This restriction allows us to present the method concisely and build the baseline for presenting the most general case of arbitrarily shifted and tilted bases. Note that such a restriction does not preclude triaxial deformations of both bases.

We begin by considering the s.p.\ wave function, used to compute properties of nuclei in the Hartree-Fock method, expanded in the three-dimensional TCHO basis,
\begin{equation}
	\Psi_\alpha(\textbf{r}\sigma)=\sum_{i=A}^{B}\sum_{\mathbf{n}=0}^{N_0}\sum_{s_z=-1/2}^{1/2}\mathbb{C}^{ \mathbf{n},i,s_z}_{\alpha}\phi_{\mathbf{n}, i}(\mathbf{r})\delta_{s_z\sigma} ,
\label{eq:eq1}
\end{equation}
where $\alpha$ is the index of a given s.p.\ state, $\mathbb{C}^{\mathbf{n},i,s_z}_{\alpha}$ are the expansion coefficients, $\mathbf{r}=(r_x,r_y,r_z)$ is the Cartesian position vector, and $\mathbf{n}=(n_x,n_y,n_z)$ represents the vector of the Cartesian HO quantum numbers. For clarity, here and below, we omit the isospin indices of wave functions and matrices. In the Cartesian representation, for the $i$-th center (denoted by $A$ or $B$), the wave function $\phi_{\mathbf{n},i}(\mathbf{r})$ is the product of the shifted and deformed one-dimensional HO basis states,
\begin{equation}
\phi_{\mathbf{n},i}(\mathbf{r}) = \varphi_{n_x,i}(r_x)\varphi_{n_y,i}(r_y)\varphi_{n_z,i}(r_z).
\label{eq:eq1a}
\end{equation}
Note that the wave functions of both centers are here represented in the common reference frame; that is, we use shifted wave functions instead of shifted reference frames.
In Eq.~(\ref{eq:eq1}), for each center, summation over vector $\mathbf{n}$ represents the sum over the HO quantum numbers suitably restricted to $M_i$ lowest HO states as defined in the OCHO case in Ref.~\cite{Dob1997-2}.
 
The one-dimensional components of $\phi_{\mathbf{n},i}(\mathbf{r})$ are defined as
\begin{align}
\varphi_{n_\mu,i}(r_\mu) &= \sqrt{\frac{b_{\mu,i}}{\sqrt{\pi}2^{n_\mu}n_\mu!}} \nonumber \\
&\quad \times H_{n_\mu}\Big(b_{\mu,i}(r_\mu-r_{\mu 0,i})\Big)e^{-\frac{1}{2}b_{\mu,i}^2(r_\mu-r_{\mu 0,i})^2},
\label{eq:eq2}
\end{align}
where $\mu=x$, $y$, or $z$ and the standard HO constants are defined as $b_{\mu, i}=\sqrt{m\omega_{\mu,i}/\hbar}$. 
Absorbing the factor $\left(\sqrt{\pi}2^{n_\mu}n_\mu!\right)^{-1/2}$ in the normalized Hermite polynomials $H_{n_\mu}^{(0)}$,
and using the dimensionless variables defined as
\begin{eqnarray}
\label{eq:eq2b1}
\xi_{\mu,i}&=&b_{\mu,i}(r_\mu-r_{\mu 0,i}), 
\end{eqnarray}
the wave functions~(\ref{eq:eq2}) take the form:
\begin{align}
\varphi_{n_\mu,i}(r_\mu) =
&\sqrt{b_{\mu,i}}H_{n_\mu}^{(0)}(\xi_{\mu,i})e^{-\frac{1}{2}\xi_{\mu,i}^2}.
\label{eq:eq2a}
\end{align}
The OCHO basis is trivially obtained by setting $\textbf{r}_{0,A}=\textbf{r}_{0,B}=0$ and $b_{\mu,A}=b_{\mu,B}$.

Even though code {\sc hfodd} takes advantage of simplex symmetry to accelerate calculations~\cite{Dob1997}, our goal is to describe the complex motion of the fission fragments, such as bending or wriggling~\cite{Bulgac2021}. Hence, the system is generally not invariant under the simplex transformation. Nevertheless, the matrix structure of coefficients $\mathbb{C}$ and of every one-body operator, $\mathbb{O}$, has the following generic form,
\begin{equation}
\mathbb{O} =
\begin{pmatrix}
O_{AA}^{++} & O_{AB}^{++} & O_{AA}^{+-} & O_{AB}^{+-} \\
O_{BA}^{++} & O_{BB}^{++} & O_{BA}^{+-} & O_{BB}^{+-} \\
O_{AA}^{-+} & O_{AB}^{-+} & O_{AA}^{--} & O_{AB}^{--} \\
O_{BA}^{-+} & O_{BB}^{-+} & O_{BA}^{--} & O_{BB}^{--}
\label{eq:eq3}
\end{pmatrix},
\end{equation}
where the superscripts represent $+i$ and $-i$ simplex.

\subsection{The generalized eigenvalue problem}

The deformed and shifted wave functions (\ref{eq:eq2}) corresponding to both centers are no longer mutually orthogonal. Hence, the HF equations represented on the non-orthogonal basis must be rewritten as
\begin{equation}
\mathbb{H}\mathbb{C}=\mathbb{N}\mathbb{C}e.
\label{eq:eq4}
\end{equation}
That is, we need to solve a generalised eigenvalue problem, which will be detailed in the next section. In equation (\ref{eq:eq4}), $\mathbb{H}$ represents the mean-field Hamiltonian matrix, $\mathbb{C}$ is the matrix of coefficients defined in Eq.~(\ref{eq:eq1}), $e$ is the diagonal matrix of the s.p.\ energies, and $\mathbb{N}$ is the norm overlap matrix, given by
\begin{equation}
\mathbb{N} =
\begin{pmatrix}
\mathbb{I}_{AA} & N_{AB} \\
N_{BA} & \mathbb{I}_{BB}
\label{eq:eq5}
\end{pmatrix},
\end{equation}
where $\mathbb{I}$ is the identity matrix and 
\begin{equation}
(N_{AB})_{\mathbf{nm}}=\int d\mathbf{r}\phi_{\mathbf{n},A}^*(\mathbf{r})\phi_{\mathbf{m},B}(\mathbf{r}).
\label{eq:eq6}
\end{equation}

To solve the HF equations using non-orthogonal bases, several strategies have been applied in the past: In the two-center shell model calculations, the basis was usually orthogonalized through the Gram-Schmidt procedure~\cite{Ong1975}, whereas in the HF calculations, a new orthogonal basis was typically defined by diagonalizing the matrix $N^\dagger N$~\cite{Berger1980}. In our case, as in the molecular-physics implementations, we chose Löwdin's canonical orthogonalization~\cite{Lowdin1950} method. In nuclear physics, Löwdin's orthogonalization has been widely used in the generator coordinate method, leading to the so-called Griffin-Hill-Wheeler equation~\cite {rin80}.

The main idea of the canonical orthogonalization is to solve the HF equations in a subspace where the eigenvalues of the norm overlap matrix~(\ref{eq:eq5}), here referred to as $\zeta$, which are smaller than a certain cutoff threshold $\zeta_{\text{cut}}$, are removed and a smaller set of orthogonal wave functions is built as
\begin{equation}
\Lambda_k(\mathbf{r})=\sum_{i=A}^B\sum_{\mathbf{n}=0}^{N_0}\frac{u_{k\mathbf{n}}}{\sqrt{\zeta_k}}\phi_{\mathbf{n},i}(\mathbf{r}) = \sum_{i=A}^B\sum_{\mathbf{n}=0}^{N_0}U_{k\mathbf{n}}\phi_{\mathbf{n},i}(\mathbf{r}),
\label{eq:eq7}
\end{equation}
where $u_{k\mathbf{n}}$ are the eigenvectors of the norm overlap matrix. Then, the non-rectangular transformation $U$ fulfils $U^\dagger\mathbb{N}U=\mathbb{I}$ and allows us to transform the generalised eigenvalue problem into the orthogonal case. 

When solving the HFB equations, one must consider both $p$-$h$ and $p$-$p$ channels. In this case, the HFB matrix in the orthogonal basis can be obtained as follows:

\begin{equation}
\mathcal{H}^{'}_{\text{HFB}} = 
\begin{pmatrix}
U & 0 \\
0 & U^*
\end{pmatrix}
\begin{pmatrix}
h - \lambda_F & \Delta \\
-\Delta^* & -h^* + \lambda_F
\end{pmatrix}
\begin{pmatrix}
U^\dagger & 0 \\
0 & U^T
\end{pmatrix},
\end{equation}
where $h$ and $\Delta$ stand for the mean field and pairing field, respectively, and $\lambda_F$ is the usual chemical potential, which ensures the correct number of particles in the system. Once the HF (HFB) equations are solved, the single (quasi)-particle wave functions can be obtained in the TCHO basis using the same transformation.

\subsection{Matrix elements of the Skyrme mean field}

One of the crucial ingredients of the local DFT is the particle density in space, which is needed to compute the matrix elements of the mean field as well as the energy of the system. The local density of nucleons reads
\begin{equation}
\rho(\mathbf{r}\sigma'\sigma)=\sum_\alpha v_\alpha^2 \Psi_\alpha(\textbf{r}\sigma')\Psi_\alpha^*(\textbf{r}\sigma),
\label{eq:eq12}
\end{equation}
where $v_\alpha^2$ is the occupation factor of the $\alpha$-th s.p.\ state. Hence, using the TCHO basis, it can be expanded as
\begin{equation}
\rho(\mathbf{r}\sigma'\sigma)=\rho_{AA}(\mathbf{r}\sigma'\sigma)+\rho_{BB}(\mathbf{r}\sigma'\sigma)+2\text{Re}\left[\rho_{AB}(\mathbf{r}\sigma'\sigma)\right].
\label{eq:eq14}
\end{equation}
Not only the local density but also all other quasi-local densities that build the Skyrme functional~\cite{Engel1975,(Per04c)} have the TCHO form of Eq.~(\ref{eq:eq14}) and the arguments presented below apply also to them.

One of the principal advantages of using the HO basis in the local DFT, which stems from the particular form of the wave functions~(\ref{eq:eq2a}), is the fact that the local densities are always in the form of the products of polynomials $W_{ij}$ and Gaussian factors~\cite{(Cop66),Dob1997}. In the TCHO basis, omitting for clarity the spin degrees of freedom this gives
\begin{equation}
\rho_{ij}(\mathbf{r})=W_{ij}(r_x,r_y,r_z)e^{-\frac{1}{2}\sum_\mu\left(\xi_{\mu,i}^2+\xi_{\mu,j}^2\right)}.
\label{eq:eq14b}
\end{equation}
The choice of the Gauss-Hermite quadrature of an appropriate order then allows for evaluating all integrals of densities {\it exactly}, that is, overall excellent stability and resilience to the rounding errors of the Gauss-Hermite quadrature allow for obtaining the results within the machine precision, usually of the order of 10$^{-15}$ for the double-precision arithmetics.

In particular, let us consider evaluating the matrix elements of a given term associated with the Skyrme interaction~\cite{Engel1975,(Per04c)}. Apart from the density-dependent term, which we discuss later, all mean fields are linear in densities and thus also have the form of the products of polynomials $G_{ij}$ and Gaussian factors
\begin{equation}
O_{ij}(\mathbf{r})=G_{ij}(r_x,r_y,r_z)e^{-\frac{1}{2}\sum_\mu\left(\xi_{\mu,i}^2+\xi_{\mu,j}^2\right)}.
\label{eq:eq14a}
\end{equation}
Then, we obtain the space part of the matrix element as
\begin{equation}
O_{\mathbf{n}i,\mathbf{m}j}(i'j')=\int_{\mathbb{R}^3} {\rm d}\mathbf{r}\, \phi_{\mathbf{n},i}^*(\mathbf{r}) O_{i'j'}\left(\mathbf{r}\right)\phi_{\mathbf{m},j}(\mathbf{r}),
\label{eq:eq15}
\end{equation}
and from Eqs.~(\ref{eq:eq1a}), (\ref{eq:eq2a}), and~(\ref{eq:eq14a}) we then have
\begin{equation}
\begin{split}
&O_{\mathbf{n}i,\mathbf{m}j}(i'j')=\\
&\int_{\mathbb{R}^3} {\rm d}\mathbf{r}\, G_{i'j'}(r_x,r_y,r_z)\prod_{\mu=x,y,z}
\sqrt{b_{\mu,i}b_{\mu,j}}
H_{n_\mu}^{(0)}(\xi_{\mu,i})\\
&H_{m_\mu}^{(0)}(\xi_{\mu,j})e^{-\frac{1}{2}\left(\xi_{\mu,i'}^2+\xi_{\mu,j'}^2+\xi_{\mu,i}^2+\xi_{\mu,j}^2\right)}.
\label{eq:eq15a}
\end{split}
\end{equation}
The integrand above contains products of four types of Gaussian factors each corresponding to either center $A$ or $B$, that is, 16 possible combinations. However, it is easy to see that only five partitions of the products of four Gaussian factors suffice, see~\ref{appendixA}.

To use the Gauss-Hermite quadrature, we need to transform every integral of Eq.~(\ref{eq:eq15a}) into the usual structure as
\begin{equation}
    \int_{-\infty}^{+\infty} f(\eta)e^{-\eta^2}d\eta = \sum_{q=1}^{N_q}\omega_q f(\eta_q),
\label{eq:eq21}
\end{equation}
where $f(\eta)$ is a polynomial, $\omega_q$ and $\eta_q$ are the weights and nodes of the quadrature, respectively, and $N_q$ is its order. For that purpose, we combine the exponents to obtain one single Gaussian, which defines the lattice of the quadrature in function of different combinations of center indices.

Owing to the properties of the Gaussians, only five different lattices are enough (see~\ref{appendixA} for full details). However, considering the structure of densities shown in Eqs.~(\ref{eq:eq14}) and~(\ref{eq:eq14b}), the number of polynomials to be evaluated is notably higher in comparison with the usual one-center bases. In the most demanding case, when pairing and density-dependent (DD) interactions are considered, up to 20 different polynomials must be computed (see~\ref{appendixC}).

At this point, we note that an analogous implementation can be used for non-co-axial bases, that is, those not only arbitrarily shifted and deformed but also arbitrarily tilted. Indeed, in this case, the s.p.\ wave functions (\ref{eq:eq2a}) depend on dimensionless variables $\xi_{\mu,i}$ given by
\begin{eqnarray}
\label{eq:eq2b3}
\xi_{\mu,i}&=&{\textstyle\sum_\nu}\,R_{\mu\nu}^ib_{\nu,i}(r_\nu-r_{\nu 0,i}), 
\end{eqnarray}
where $R_{\mu\nu}^i$ are the 3$\times$3 orthogonal rotation matrices for centers A and B. 

Even for non-co-axial bases, the integration lattices remain the same, as the exponent of the common Gaussian is invariant under rotations. Thus, the expressions in this section and the appendices are still valid. The use of non-co-axial bases will be discussed in a forthcoming publication~\cite{(San25)}.

Regarding the inclusion of pairing correlations, instead of using the antisymmetric pairing tensor $\kappa$, we use the pairing density matrix~\cite{Dob84}, defined as:
\begin{equation}
    \tilde{\rho}(\mathbf{r}\sigma,\mathbf{r}'\sigma')=-2\sigma'\braket{\Psi|a_{\mathbf{r}'-\sigma}a_{\mathbf{r}\sigma}|\Psi},
\end{equation}
which allows us to apply the same method to the matrix elements of the $p$-$p$ channel. However, it requires evaluating additional polynomials on the quadrature lattices (see~\ref{appendixC}), which leads to a substantial increase in computational time.  Thus, we have opted not to include pairing correlations in our Proof-of-Principle calculations, as they are primarily intended to demonstrate the method's capabilities rather than to provide realistic results, which will be the focus of future publications.  

\subsection{Coulomb interaction in the TCHO basis}

Proper treatment of the Coulomb interaction is crucial when describing fission or any reaction involving two nuclei. The Coulomb force acts along the whole fission path, affecting the neck formation, the interaction between the pre-fragments, and the evolution of each fragment after scission~\cite{Bender2020}. The advantage of using the TCHO basis lies in describing all stages of the fission process by employing different shifts of the bases. Hence, the TCHO method implemented here allows us to describe the effects of the Coulomb interactions not only in each fragment but also between them on the way to and after the scission. This is vital to describe, among other observables, the total kinetic energy of the reaction~\cite{Schunck2016}. Our TCHO implementation of the exact Coulomb exchange effects is critical, as its consequences for neck formation and fragment distributions have never been considered.

The Coulomb interaction can be represented as
\begin{equation}
\hat{V}(\mathbf{r}_1,\mathbf{r}_2)=\frac{e^2}{|\mathbf{r}_1-\mathbf{r}_2|}
\hat{\sigma}_0^{(1)}\hat{\sigma}_0^{(2)}\delta_{\tau,p}^{(1)}\delta_{\tau,p}^{(2)}\left(1-\hat{P}^{\sigma}\hat{P}^{\tau}\hat{P}^{\mathbf{r}}\right),
\label{eq:Coul}
\end{equation}
where indices 1 and 2 pertain to coordinates of two interacting particles, $\hat{\sigma}_0$ are the 2$\times$2 diagonal spin matrices, and $\hat{P}^{\sigma}$, $\hat{P}^{\tau}$, and $\hat{P}^{\mathbf{r}}$ are the standard spin, isospin, and space exchange operators, respectively. The direct, exchange, and pairing matrix elements of the Coulomb interaction can be effectively treated by exchanging the space indices of the direct term, see Ref.~\cite{(Sch17d)}. Therefore, below, we discuss only the space part of the direct term, that is,
\begin{equation}
\begin{split}
&\braket{\mathbf{n},{i};\mathbf{m},{j}|V^{\mathrm{dir}}(\mathbf{r}_1,\mathbf{r}_2)|\mathbf{n}',{i'};\mathbf{m}',{j'}}=\\
& e^2
\iint d\mathbf{r}_1 d\mathbf{r}_2  \phi_{\mathbf{n},i}^*(\mathbf{r}_1)\phi_{\mathbf{m},j}^*(\mathbf{r}_2)\frac{1}{|\mathbf{r}_1-\mathbf{r}_2|}\phi_{\mathbf{n}',i'}(\mathbf{r}_1)\phi_{\mathbf{m}',j'}(\mathbf{r}_2)
\label{eq:eq62}
\end{split}
\end{equation}
To evaluate this matrix element by employing the Gauss-Hermite quadratures again, we use the method introduced by Girod and Grammaticos~\cite{(Gir83a)} and later implemented in numerous codes~\cite{(Dob96b), Dob2009,(Sto13),(Mar22),(Zur24)}. The method relies on approximating the Coulomb potential by a sum of Gaussians,
\begin{equation}
    \frac{1}{|\mathbf{r}_1-\mathbf{r}_2|}\simeq\sum_{\gamma=1}^{N_C}A_\gamma e^{-a_\gamma(\mathbf{r}_1-\mathbf{r}_2)^2},
    \label{eq:eq63}
\end{equation}
where at any order $N_C$, the strengths $A_\gamma$ and widths $a_\gamma$ can be evaluated by simple algebraic expressions. The key point now is that the matrix elements become separable; that is, they are products of matrix elements separately evaluated in each Cartesian direction,
\begin{equation}\label{eq:eq64}
\begin{split}
    &\braket{\mathbf{n},{i};\mathbf{m},{j}|V^{\mathrm{dir}}(\mathbf{r}_1,\mathbf{r}_2)|\mathbf{n}',{i'};\mathbf{m}',{j'}} =\\
    &\hspace{20mm}e^2\sum_{\gamma=1}^{N_C}A_\gamma\prod_{\mu=x,y,z} v^\gamma_{\mathbf{n}_\mu,{i};\mathbf{m}_\mu,{j};\mathbf{n}'_\mu,{i'};\mathbf{m}'_\mu,{j'}},
\end{split}
\end{equation}
where
\begin{equation}\label{eq:eq65}
\begin{split}
    &v^\gamma_{\mathbf{n}_\mu,{i};\mathbf{m}_\mu,{j};\mathbf{n}'_\mu,{i'};\mathbf{m}'_\mu,{j'}} = 
    \iint dr_{1\mu} dr_{2\mu} H_{n_\mu}^{(0)}(\xi_{1\mu,i}) \\
    &~~\times H_{m_\mu}^{(0)}(\xi_{2\mu,j})H_{n'_\mu}^{(0)}(\xi_{1\mu,i'})H_{m'_\mu}^{(0)}(\xi_{2\mu,j'}) \\
    &~~\times  e^{-\frac{1}{2}(\xi_{1\mu,i}^2+\xi_{1\mu,i'}^2)} e^{-\frac{1}{2}(\xi_{2\mu,j}^2+\xi_{2\mu,j'}^2)}e^{-a_\gamma(r_{1\mu}-r_{2\mu})^2}.
\end{split}
\end{equation}
The last step to compute this integral by the Gauss-Hermite quadrature is to express the product of Gaussians in terms of a Gaussian of the quadratic form in variables $\xi_1$ and $\xi_2$, which is presented in detail in~\ref{appendixD}.

From this point on, the calculation of Coulomb energies and mean fields in the direct, exchange, and pairing channels proceed in close analogy to those implemented for any other finite range interaction, such as Yukawa or Gogny force, which was described in Refs.~\cite{Dob2009,(Sch17d)},

\begin{figure}[htb]
\begin{center}
\includegraphics[width=\linewidth]{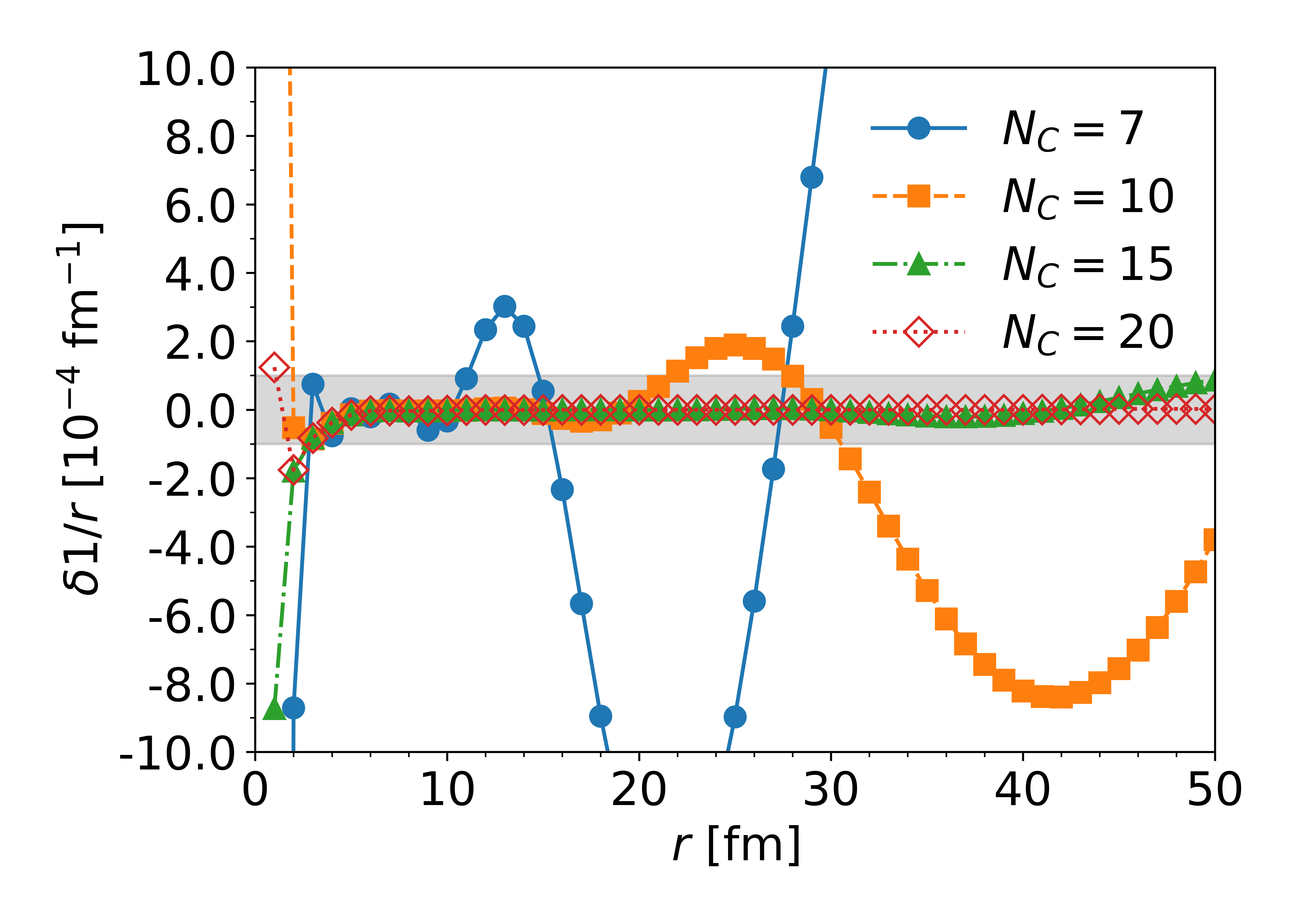}
\end{center}
\caption{\label{gaussian_test} (Color online) Difference between the exact form factor $1/r$ and the expansion~(\protect\ref{eq:eq63}) computed for different numbers of Gaussians $N_C$. The gray shaded area shows a precision of 10$^{-4}$ fm$^{-1}$ or better.}
\end{figure}

Even though for the one-center calculations, the number of Gaussians $N_C$ that ensure the precision of the $1/r$ expansion within the nuclear volume is less than 10~\cite{Dob1997}, for two fragments separated by large distances this cannot be enough. This is illustrated in Fig.~\ref{gaussian_test}, where we show the Coulomb form factor $1/r$ compared with expansion~(\ref{eq:eq63}) for different numbers of Gaussians $N_C$. We see that expansion on $N_C=20$ Gaussians is sufficiently precise for distances up to about 50\,fm. As the computation time scales linearly with $N_C$, higher values can be easily accommodated if necessary.

\section{Proof of Principle I: testing accuracy in the case of $^8\text{Be}$}\label{section3}

\subsection{Ground state energies: one-center vs. two-center calculation}

The ultimate goal of the TCHO solver is to describe the fission path in a more accurate and, if possible, easier way. For that, we need to tell the initial nucleus before the reaction starts. Here, as proof for validating the method, we have chosen the case of $^8$Be. Its well-known $\alpha+\alpha$ cluster structure is ideal for putting the TCHO formalism to the test.

First, we compare the OCHO and TCHO ground-state results for different numbers of shells included in the calculations. For the OCHO calculations, we used a deformed basis adapted to the deformation of the ground state energy: for $Q_{20}\equiv\braket{\hat{Q}_{20}}\approx0.5$ b we have $\hbar\omega_x=\hbar\omega_y=31.94$ MeV and $\hbar\omega_z=14.59$ MeV. On the other hand, for the two-center method, we chose both bases adapted to the properties of spherical $^4$He, $\hbar\omega_x=\hbar\omega_y=\hbar\omega_z=30.99$ MeV. Furthermore, as the average nuclear radius of $^4$He is around 1.90 fm (as obtained from the HF OCHO calculation), we set the two centers of the basis at $z_A=-2$ fm and $z_B=2$ fm, with a basis cutoff of $\zeta_{\text{cut}}=10^{-4}$. The Skyrme functional used is UNEDF1~\cite{UNEDF1}, which will be used in realistic fission calculations.

\begin{figure}[ht]
\begin{center}
\includegraphics[width=\linewidth]{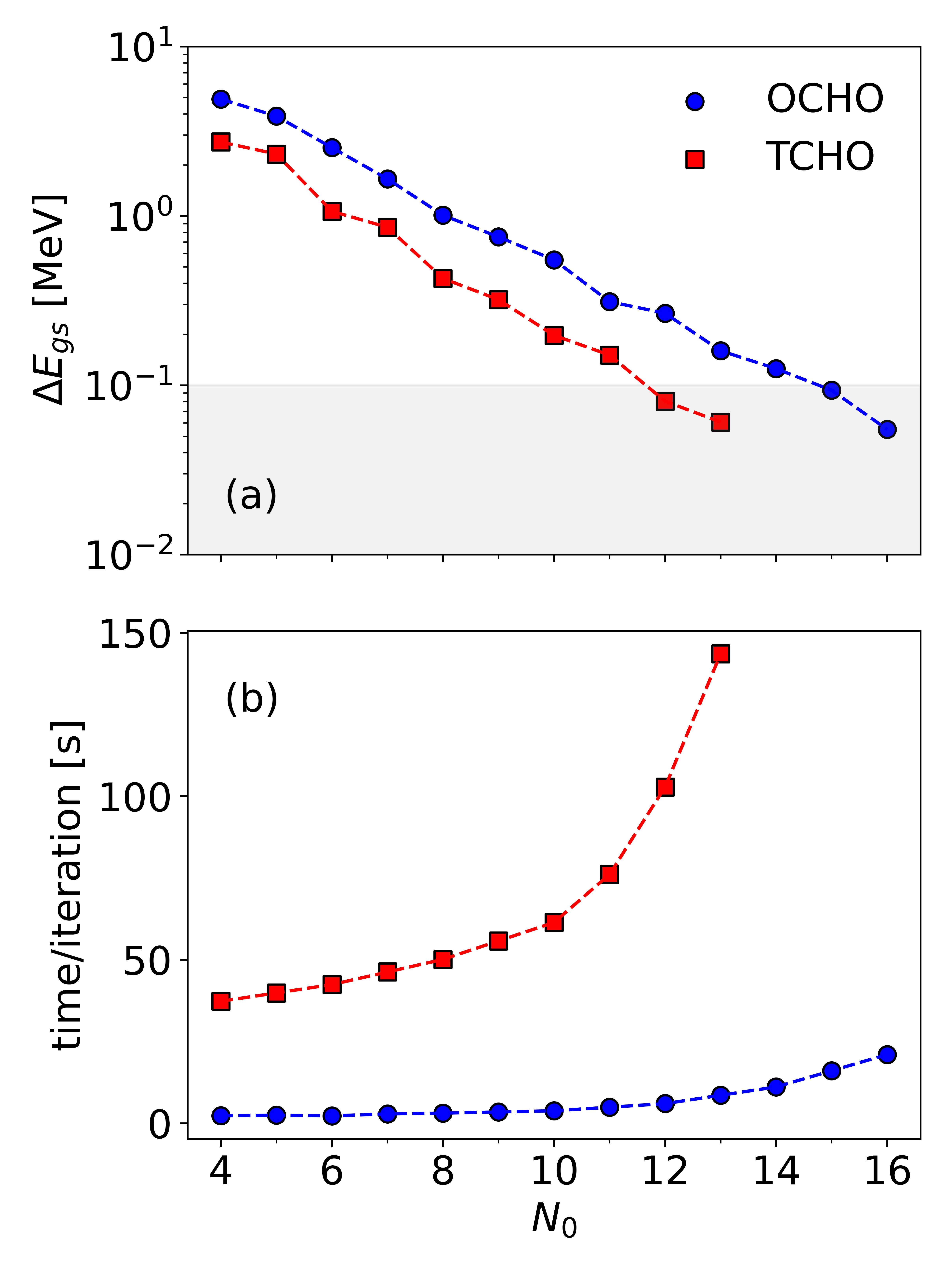}
\end{center}
\caption{\label{ocvstc} (Color online) (a) Deviation of the $^8$Be ground state energy from the asymptotic value ($-41.10$ MeV) as a function of the number of HO shells, calculated using the OCHO and TCHO bases. The gray-shaded region represents the area where the precision is equal to or better than 100 keV. (b) CPU time per iteration for the single-core version of {\sc hfodd} using the OCHO and TCHO bases (tested on the University of York Viking cluster, AMD EPYC 7643 @ 2.30 GHz).}

\end{figure}

In Fig. \ref{ocvstc}(a), we compare the deviation from the asymptotic ground-state energy for both OCHO and TCHO calculations. The asymptotic value was obtained by extrapolating to $N_0\rightarrow \infty$ the energies derived from the standard one-center basis expansion. Our results show that the TCHO basis consistently yields a lower deviation for a given number of shells, $N_0$, demonstrating a better variational approximation of the $^8$Be wave function. However, as illustrated in Fig. \ref{ocvstc}(b), this increased accuracy comes with a higher computational cost. This is because not only the Hamiltonian matrix elements for each center, but also those between two different centers, need to be calculated. Fortunately, optimizing the basis parameters can limit the number of shells to 12 while maintaining the typical precision expected in DFT calculations. This perfectly agrees with the approach of practitioners of the Gogny model, who— even for heavy fissioning systems—only include up to $N_0=11$ shells~\cite{Lemaitre2018}. However, it is important to recall that this expansion aims to compete with the more demanding OCHO bases needed to describe extremely elongated shapes, rather than configurations near the ground state. In such cases, one-center bases may require more than 30 shells~\cite{Schunck2014,Schunck2015}, making the computational burden closer to our approach. Moreover, to study the neck formation or post-scission evolution, TCHO can be important, even if it is time-consuming. Additionally, we also achieve the additional advantage of including triaxiality at the exact computational cost.

\subsection{Coulomb interaction between $\alpha$ particles: a test of the numerical precision.}

\begin{figure}[!htb]
\begin{center}
\includegraphics[width=\linewidth]{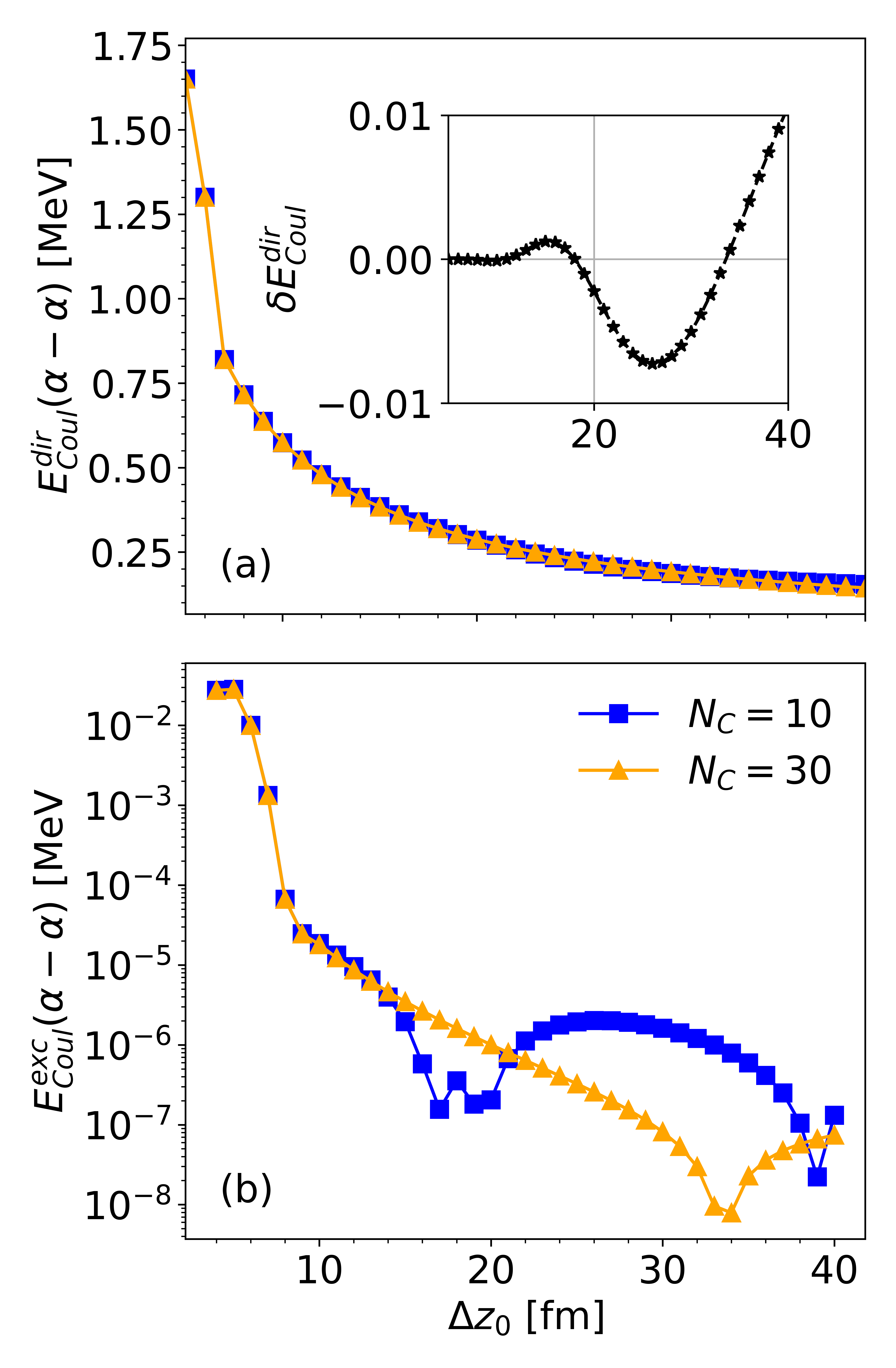}
\end{center}
\caption{\label{couldir_exc} (Color online) (a) The direct part of the Coulomb interaction between $\alpha$ particles within $^{8}$Be and after scission. The TCHO basis was used for $N_0=4$ shells, and SV$_\text{T}$~\cite{svtint} Skyrme functional. The inset shows the energy difference between the two curves $N_C=10,30$. (b) Same as (a) but in logarithmic scale for the exchange term.}
\end{figure}

Fig.~\ref{couldir_exc}(a) displays the direct part of the Coulomb interaction between the $\alpha$ particles clustering in $^8$Be for two different numbers of Gaussians as a function of the separation between the bases. We use $N_C=10$ as the standard number of Gaussians for separations up to 20 fm, while $N_C=30$ provides the reference values for highly precise Coulomb energies, accurate to within $10^{-7}$ MeV.

As seen in the figure, the differences between using 10 or 30 Gaussians are not noticeable on the plot scale until the curves diverge at 30 fm. In the TCHO method, the separation between the two fragments becomes apparent, as the Coulomb interaction then follows that of two point-like charges. Specifically, this occurs at a separation of \( \Delta z_0 \approx 7.5 \) fm, corresponding to a total quadrupole moment of $Q_{20} \approx 9$ b. Beyond this point, the electric repulsion decreases as $1/r$, ultimately converging to twice the direct Coulomb interaction between the $^4$He fragments. The structure of two separated well-defined fragments can also be seen in Fig.~\ref{couldir_exc}(b), where the exchange part of the Coulomb interaction is shown.  Once the fragments separate, the relative exchange interaction energy rapidly reaches zero, showing an exponential decrease that can only be appreciated in the logarithmic scale. This exponential behavior is up to 15\,fm well reproduced even for 10 Gaussians and up to 30\,fm for 30 Gaussians.

In the forthcoming publications, we will systematically study the impact of the exact treatment of the Coulomb exchange on neck formation and fragment distributions. Indeed, in the region of low density, the Slater approximation~\cite{(Sla51a)} of the Coulomb exchange is not justified. 

\section{Proof of Principle II: $^{24}\text{Mg}\rightarrow$ $^{12}$C+$^{12}$C}\label{section4}

As the second Proof of Principle, we analyze the symmetric fission channel in $^{24}$Mg. On the one hand, this is a well-tested reaction in the two-center formalism~\cite{Berger1980}. On the other hand, as Berger and Gogny showed, 6 HO shells are enough to describe the problem to avoid the usual DFT computational burden of a heavy nucleus. 

\subsection{Description of the $^{24}$Mg ground state}

To benchmark the results for the ground state of $^{24}$Mg, we performed the OCHO calculation using a deformed basis adapted to the ground state quadrupole moment of around 1.1\,b. In the direction of the $z$-axis, we included 17, whereas in the perpendicular directions 12 HO shells. In the case of the TCHO calculation, we kept the same strategy as in the $^{8}$Be test, adapting both bases to the properties of the spherical $^{12}$C and including up to 10 HO shells. In both calculations, we included 572 HO states (twice the size of the spherical $N_0=10$ basis). For the basis cutoff parameter in the norm eigenvalues set to $\zeta_{\text{cut}}=10^{-4}$, the TCHO method discarded a few tens of states due to the basis non-orthogonality; see section ~\ref{section2a}.

\begin{figure}[htbp]
\begin{center}
\includegraphics[width=\linewidth]{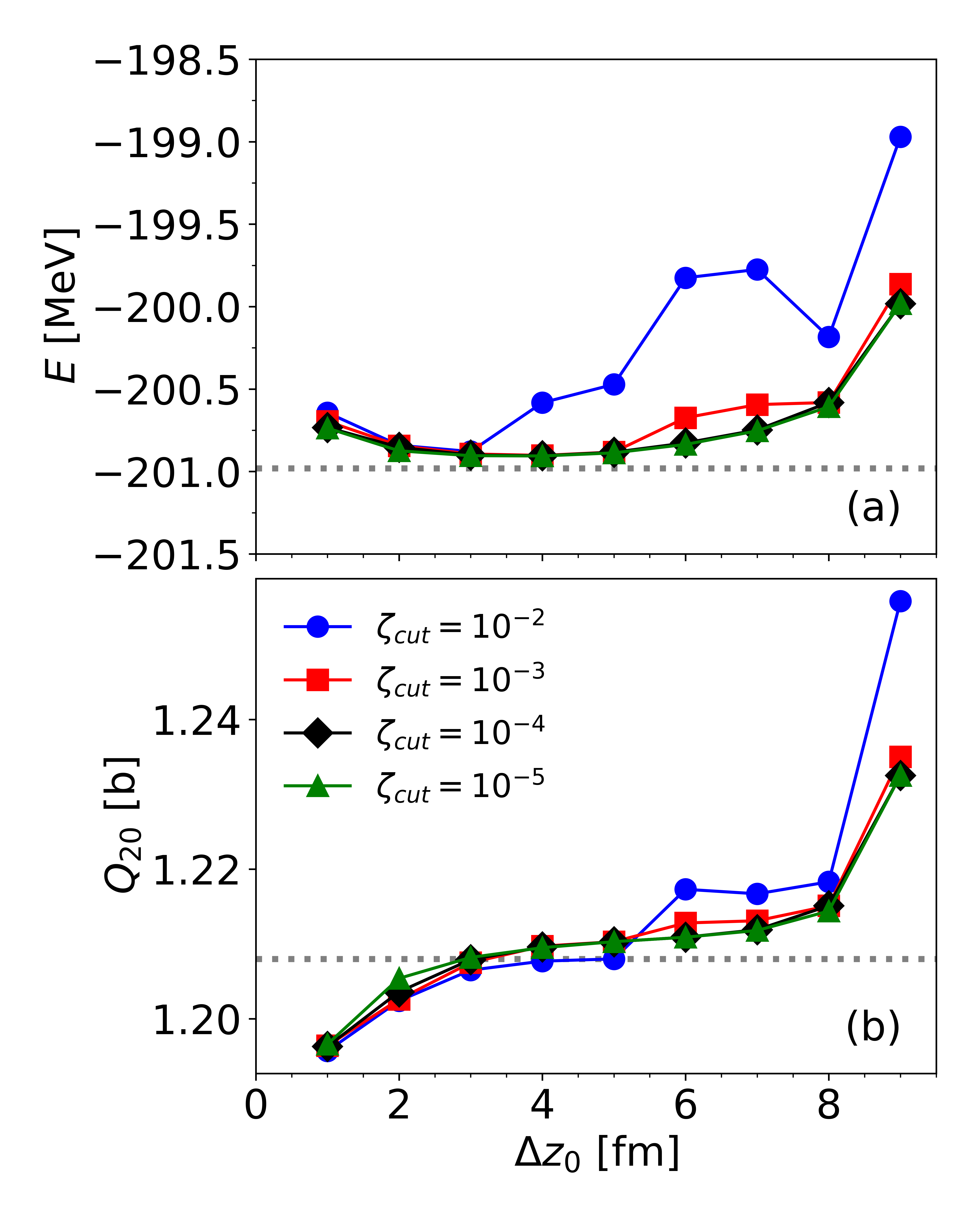}
\end{center}
\caption{\label{cutoff_svt} (Color online) Ground state energy (a) and quadrupole moment (b) of $^{24}$Mg in function of the separation of the centers for different cutoffs $\zeta_{\text{cut}}$ in the norm eigenvalues. The Coulomb interaction and pairing correlations were neglected. The grey dashed line indicates the results obtained using the OCHO basis.}
\end{figure}

To study the dependence on the center separation, we analyzed different values in the range between 0 and 9\,fm, as larger values than those lead to two completely separated fragments. We also varied the cutoff in the norm overlap $\zeta_{\text{cut}}$ for completeness. In Fig.~\ref{cutoff_svt}, we show the ground-state energies and quadrupole moments as functions of the separation of the basis for different cutoffs $\zeta_{\text{cut}}$. To avoid numerical instabilities, we used the density-independent Skyrme functional SV$_\text{T}$~\cite{svtint}; however, the same behavior was found for other Skyrme parametrizations.

\sloppy
We found that the best TCHO description of the ground-state energy was given by the separation of 4\,fm between centers, which was just 75\,keV above the OCHO result, including up to 17 HO shells. Moreover, from Fig.~\ref{cutoff_svt}, we learned that a cutoff of $10^{-4}$ is sufficient to include the relevant number of states for an accurate description. 
Related to the deformation of the ground state, we saw a weak increase of the average value of $\braket{\hat{Q}_{20}}$ in function of the separation of the centers. However, these variations represent less than 5\%.

\sloppy
Using the optimum TCHO basis, we performed a more realistic calculation using the SkM*~\cite{skmint} interaction, including the full Coulomb interaction. Some relevant quantities are summarized in Table~\ref{table_24Mg_gs}, in which we compare the results obtained for the OCHO and TCHO bases. In either case, we did not conserve simplex, signature, or parity symmetry.  In the OCHO calculation, we enforced the axial symmetry of the solution. In contrast, in the TCHO calculation, this symmetry was rapidly fixed self-consistently due to the separation of the bases along the $z$-axis. Despite the smaller number of shells included in the TCHO calculation, we observe no significant differences in any of the results, so the TCHO method captures the ground-state properties of $^{24}$Mg perfectly well. 

\begin{table}[h!]
\centering
\begin{tabular}{||c c c||} 
 \hline
  & def. OCHO & sph. TCHO \\ [0.5ex] 
 \hline\hline
 $E_{\mathrm{kin}}$ (MeV) & 387.09 & 388.15 \\ 
 $\sum E_{\mathrm{s.p.}}$ (MeV) & -492.36 & -492.44 \\
 $E_{\mathrm{Coul}}$ (MeV) & 27.90 &  27.93 \\
 $E_{\mathrm{s-o}}$ (MeV) & -22.35 & -22.41 \\
 $E_{\mathrm{Sky}}$ (MeV) & -594.47 & -595.47 \\
 $E_{\mathrm{g.s}}$ (MeV) & -179.48 & -179.38 \\
 $\braket{\hat{Q}_{20}}$ (b) & 1.17 & 1.17 \\
 $\braket{\hat{Q}_{40}}$ (b$^2$) & 0.012 & 0.012 \\ [1ex] 
 \hline
\end{tabular}
\caption{Results of HF calculations performed for $^{24}$Mg using the deformed (def.) OCHO and spherical (sph.) TCHO bases for the Skyrme functional SkM*. The rows of the Table show the kinetic energy, the sum of the single-particle energies, the Coulomb, spin-orbit, and Skyrme functional energies, the ground-state total energy, and axial quadrupole and hexadecapole moments.}
\label{table_24Mg_gs}
\end{table}

\subsection{Symmetric fission of $^{24}$Mg}

The symmetric fission process is primarily driven by an increasing quadrupole deformation, which gradually leads to the emergence of distinct fragments. One might be tempted to vary the distance between fragment centers as an alternative to increasing $Q_{20}$. However, we found that varying only $\Delta z_0$ leads to two distinct outcomes: for small distances, the calculation converged to the ground state, while for larger separations, it either returns to the ground state or produces two well-separated fragments. In other words, varying $\Delta z_0$ alone cannot capture the full range of intermediate shapes--from the ground state to post-scission. Yet, when using the quadrupole deformation to construct a one-dimensional fission path, a key question arises: How should the distance between centers be selected? In our case, based on the nuclear radius of Mg at different deformations, we adopted the following simple prescription:

\begin{enumerate}[(i)]
\item Ground state: $Q_{20} \leq 2$ b. $\Delta z_0$ = 4 fm.
\item Small deformations up to the development of a reasonably populated neck: 2 b $< Q_{20} \leq 5$ b. $\Delta z_0$ = 6 fm.
\item Larger deformations where the neck becomes thinner and approaches scission: 5 b $ < Q_{20} \leq 9$ b. $\Delta z_0$ = 10 fm.
\item Two fully separated fragments: $Q_{20} > 9$ b. We varied $\Delta z_0$ gradually, from 10 to 20 fm. 
\end{enumerate}

Based on the weak dependence of the total energy on the separation parameter, as shown in Fig.~\ref{cutoff_svt}, the variations observed for different values are expected to be on the order of a few keV. Recent studies have also reported this behavior using axial two-center states within covariant DFT\cite{Li2023}.

\begin{figure}[ht]
\begin{center}
\includegraphics[width=\linewidth]{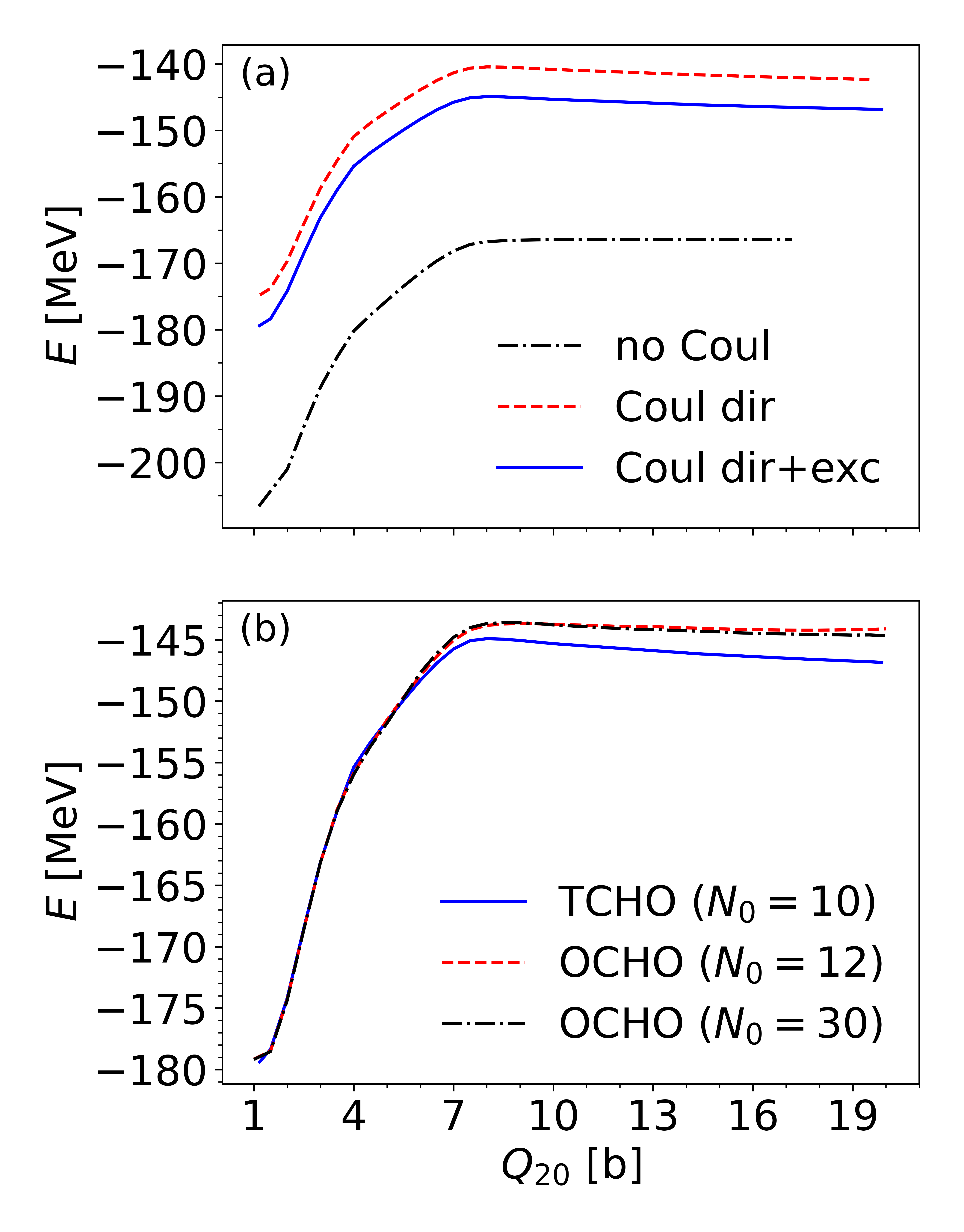}
\end{center}
\caption{\label{symfis24mg} (Color online) Total energies of $^{24}$Mg in function of the quadrupole moment, determined for the Skyrme functional SkM*. Panel (a) shows the TCHO results obtained for the $N_0=10$ spherical shells with or without Coulomb interaction. Panel (b) compares the results obtained in TCHO ($N_0=10$) and OCHO ($N_0=12$ or 30 with 1000 HO states), both with the full Coulomb interaction included.
}
\end{figure}

\begin{figure}[ht]
\begin{center}
\includegraphics[width=\linewidth]{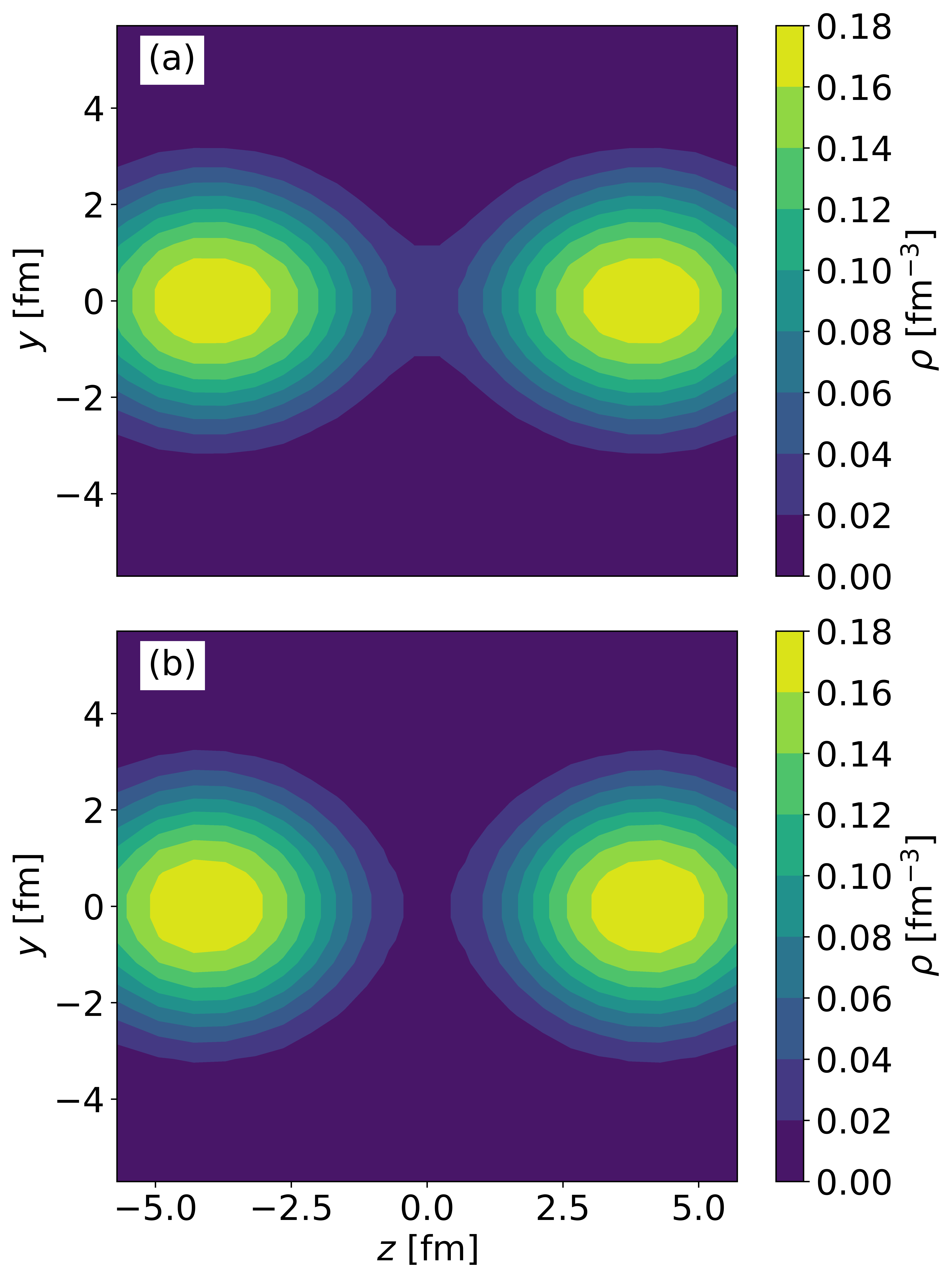}
\end{center}
\caption{\label{new_dens} (Color online) The $^{24}$Mg total density contour plots for $Q_{20}=7.5$\,b (a) and $Q_{20}=8.0$\,b\, (b) calculated using the TCHO method.}
\end{figure}

\sloppy
In Fig.~\ref{symfis24mg}(a), we show the TCHO results separately for the cases without Coulomb interaction, with the Coulomb direct term only, and with the full direct+exchange terms included. Although the no-Coulomb results are not physically realistic, they serve to test the method's ability to describe the fissioning system: from $Q_{20}\simeq8$ b onward, the curve stabilizes at the energy of two spherical $^{12}$C fragments, computed using the same basis parameters as in the TCHO $^{24}$Mg calculation. When the Coulomb interaction is included, we observe a small barrier at $Q_{20}\simeq7$ b for both the direct and direct+exchange curves. Furthermore, the energy difference between the direct and direct+exchange Coulomb curves remains almost constant, with only a slight increase starting at $Q_{20}\approx 5$ b. This suggests that the exchange term has minimal impact on this particular reaction, as it only shifts the energy curves downward. Beyond $Q_{20}\simeq8$\ b, the electrostatic repulsion starts behaving like $1/r$, corresponding to the two point-like charged separated particles. Hence, we can identify the scission occurs in the region around 8\,b. This is also confirmed by the particle density distribution, shown in Fig.~\ref{new_dens} for $Q_{20}=$7.5 and 8.0\,b. We observe that the low-density neck vanishes at 8.0\,b, resulting in two independent fragments with nearly spherical shapes.

In Fig.~\ref{symfis24mg}(b), we present the TCHO results alongside the OCHO results, both including the full Coulomb interaction. For the OCHO basis, the energy was not accurately reproduced at very high quadrupole moments, even with the inclusion of 30 HO shells and the additional constraints previously discussed.

Another important aspect is that the TCHO results were obtained without any additional constraints aside from the quadrupole moment. In contrast, we needed to impose a constraint on the hexadecapole moment, $Q_{40}$, for the OCHO case to ensure continuity and, in some instances, even convergence. This is due to the variational nature of the problem, where the nucleus may reorganize its collective degrees of freedom when converging to the minimum energy solution, potentially leading to discontinuities. On the other hand, the structure of the pre-fragments in the TCHO basis naturally reinforces the desired shape, providing extra stability for different $Q_{40}$ solutions existing at a given $Q_{20}$.

\begin{figure}[ht]
\begin{center}
\includegraphics[width=\linewidth]{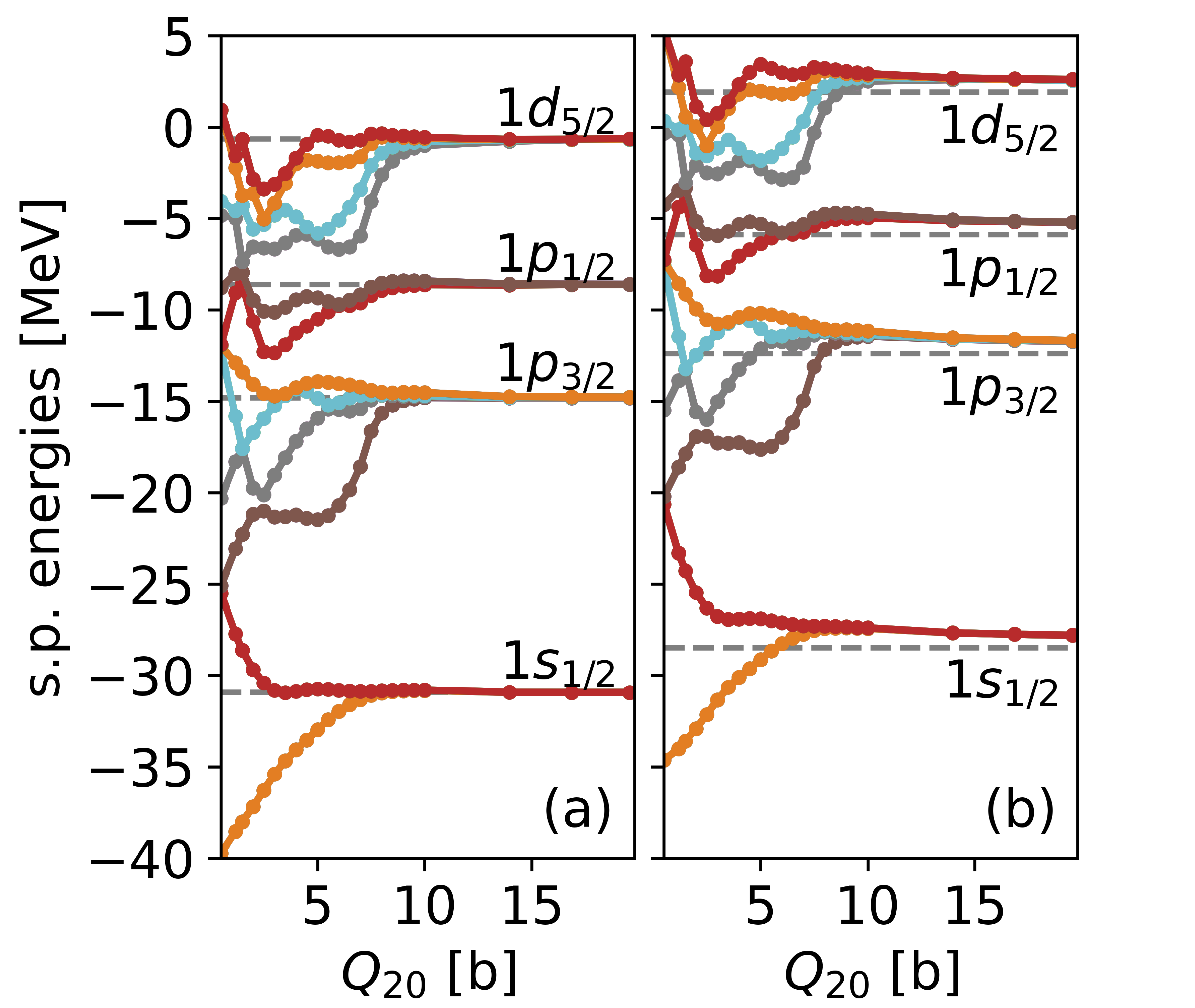}
\end{center}
\caption{\label{spener} (Color online) Twelve lowest Kramers-degenerate s.p.\ neutron (a) and proton (b) energies of $^{24}$Mg in function of $Q_{20}$ calculated in the TCHO basis (two highest unbound states that end up in the 1d5/2 orbitals of the fragments are not shown). The dashed lines show the OCHO energies of $^{12}$C calculated with the same basis parameters as those used in $^{24}$Mg for each TCHO center. For easier identification, we used the spherical HO quantum numbers of the fragments.}
\end{figure}

In Fig.~\ref{spener}, we show the $^{24}$Mg twelve lowest proton and neutron s.p.\ energies calculated in the TCHO basis, in function of the quadrupole moment, cf.\ Ref.~\cite{Berger1980}. To better appreciate the evolution of the levels, we included the values for $Q_{20}=0.5$\,b computed with $\Delta z_0=0$\,fm. This scenario is equivalent to using the OCHO basis, as the frequencies used in both centers are the same. Even though the lines connect the orbitals ordered by their energies, we can see a high degree of crossing for deformations smaller than 5\,b. Beyond that point, most of the s.p.\ states can be easily identified until they become (asymptotically) degenerate due to the structure of two separate $^{12}$C fragments (whose states are represented by the dashed lines). In the case of the proton energies, shown in panel (b), the match is imperfect due to the Coulomb repulsion between fragments, which is still noticeable after scission. Our results agree with those obtained with the two-center shell model~\cite{Chandra78}, showing the equivalence between exploring parameters $\Delta z_0 $ and $Q_{20}$ in such a symmetric case.

\begin{figure}[htb]
\begin{center}
\includegraphics[width=\linewidth]{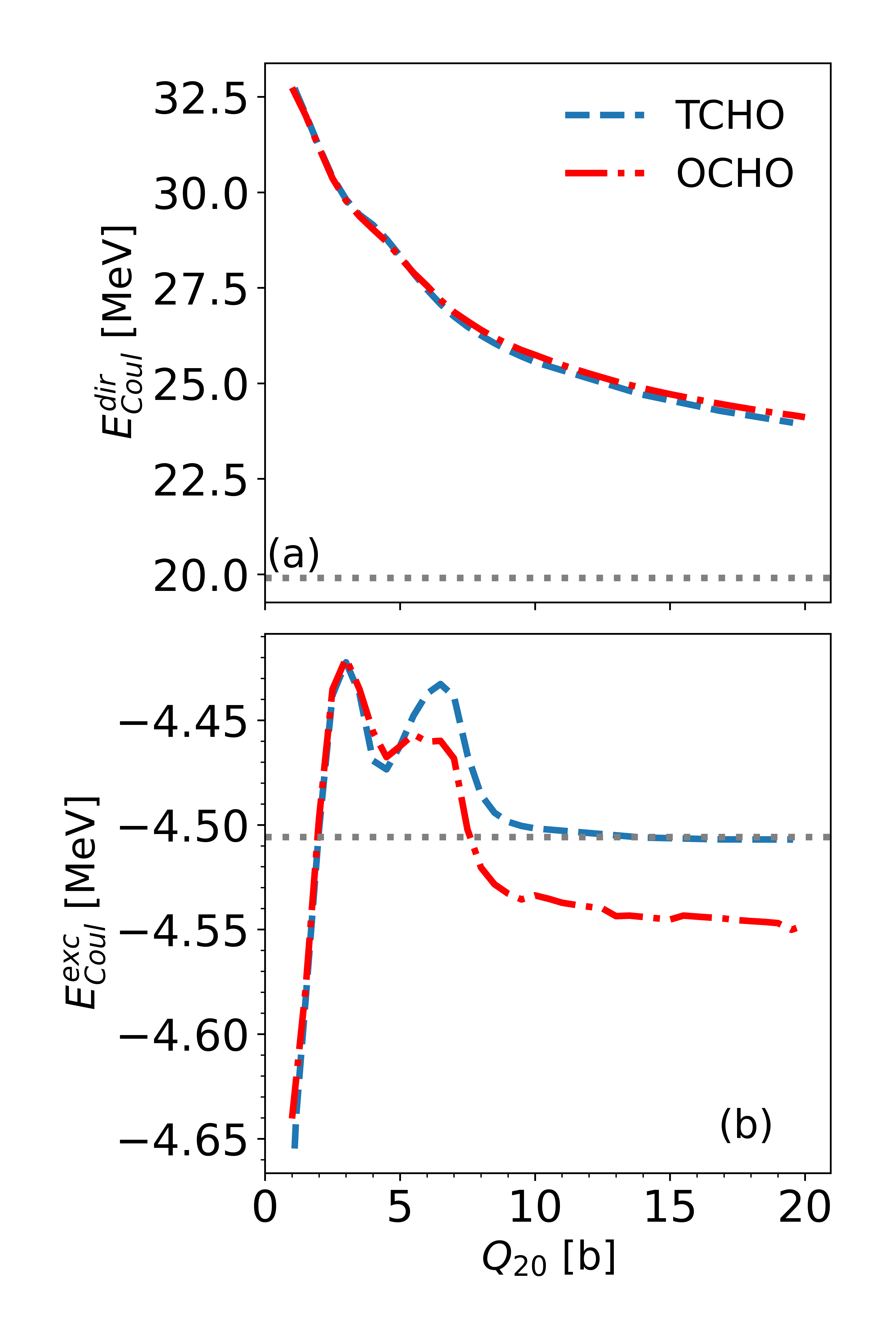}
\end{center}
\caption{\label{24mg_coulcom} (Color online) (a) Direct and (b) exchange terms of the Coulomb interaction in function of $Q_{20}$ computed with OCHO and TCHO bases and 10 Gaussians to approximate the form factor. The dotted lines represent the value of twice the interaction within the $^{12}$C nucleus, computed with 10 HO shells.}
\end{figure}

Finally, related to the Coulomb interaction, we found that both direct and exchange terms present the same behavior when OCHO or TCHO basis is used. Due to the maximum distance between fragments, given by the largest quadrupole deformation considered, we approximated the form factor by $N_C=10$ Gaussians. For the direct part, shown in Fig.~\ref{24mg_coulcom}(a), the TCHO and OCHO curves exhibit similar behavior, although slight differences are noticeable near the scission point, around 7.5-8.0\,b. For the exchange part, Fig.~\ref{24mg_coulcom}(b), due to the scale in which it is represented, the differences appear more clearly, with the TCHO results converging to the doubled exchange energies of the fragments and the OCHO results missing about 50\,keV asymptotically.

\section{Conclusions and future work}\label{section5}

In this work, we presented for the first time an unabridged 3D two-center harmonic oscillator basis method implemented in a sef-consistent solver that allows for the breaking of symmetries such as time-reversal, parity, simplex, and axiality. This initial publication aimed to establish and apply the formalism to two simple systems to test the method's accuracy. In both $^8$Be and $^{24}$Mg, we discovered that the post-scission configurations could be perfectly described due to the ability to choose the separation of the centers. We also demonstrated that the electrostatic direct and exchange interaction is well reproduced, provided the Coulomb form factor is approximated accurately enough.

Regarding the pre-scission states, we observed that the TCHO basis method can accurately describe cluster configurations, even for a few HO shells in both bases. However, one must select a reasonable center separation to characterize different deformed states. Fortunately, the range of these separations is quite broad, and the penalty of not choosing the optimum value is not significant enough to be concerning. Comparing the same symmetric fission channel in $^{24}$Mg, using OCHO and TCHO bases, we found that the latter produces the same results. Considering that the number of states in the TCHO method was relatively small and we did not include constraints on the neck density or higher-order multipole moments, the two-center method provides a more straightforward and natural tool to describe complex phenomena such as fission. 

Now that the foundational framework of the method is established, the forthcoming publication will focus on analyzing the tilting of the two-center bases and its impact on both scission and post-scission configurations of a heavy nucleus, such as $^{240}$Pu. Specifically, the role of the exchange term in the Coulomb interaction will be examined in realistic fission paths. In addition, a newly developed technique will enable us to track the evolution of both mass and deformation of the resulting fission fragments. However, considerable work remains to be done. To properly analyze the generation and evolution of angular momenta in the fission fragments, allowing the bases to have distinct orientations will be essential. Furthermore, to rigorously test the adiabatic approximation, we must permit the two HO bases—along with their separations and orientations—to evolve over time~\cite{Dob2019}.

\section*{Acknowledgments}
This work was partially supported by the STFC Grant No. ST/W005832/1, ST/P003885/1 and ST/V001035/1 and by the Fonds de la Recherche Scientifique - FNRS and the Fonds Wetenschappelijk Onderzoek - Vlaanderen (FWO) under the EOS Project No O000422. We acknowledge the CSC-IT Center for Science Ltd., Finland, for allocating computational resources. This project was partly undertaken on the Viking Cluster, a high-performance computing facility provided by the University of York. We are grateful for computational support from the University of York High-Performance Computing service, Viking, and the Research Computing team.

\appendix

\section{Integration of the mean-field matrix elements and energy density}
\label{appendixA}

From Eq.~(\ref{eq:eq15}), we see that the general matrix element between the HO states belonging to centers $i,j$ is computed as

\begin{equation}
O_{\mathbf{n} i,\mathbf{m} j}=\sum_{i'j'=A}^B\int_{\mathbb{R}^3} {d \mathbf{r}}\, \phi_{\mathbf{n},i}^*(\mathbf{r}) O_{i'j'}\left(\mathbf{r}\right)\phi_{\mathbf{n},j}(\mathbf{r}).
\label{eq:a1}
\end{equation}

The problem, as stated in section \ref{section2}, is to evaluate this integral numerically using the Gauss-Hermite quadratures. Considering the structure of the operators, each integral is the sum of three terms (only one off-diagonal block is enough due to hermiticity):

\begin{equation}
\begin{split}
&O_{\mathbf{n} A,\mathbf{m} A}=b_{x,A}b_{y,A}b_{z,A}\cdot\\
&\int_{\mathbb{R}^3} d\mathbf{r} \left[G_{AA}(\mathbf{r})\prod_\mu H_{n_\mu}^{(0)}(\xi_{\mu,A})H_{m_\mu}^{(0)}(\xi_{\mu,A})e^{-2\xi_{\mu,A}^2}+\right.\\
&\left.G_{BB}(\mathbf{r})\prod_\mu H_{n_\mu}^{(0)}(\xi_{\mu,A})H_{m_\mu}^{(0)}(\xi_{\mu,A})e^{-(\xi_{\mu,A}^2+\xi_{\mu,B}^2)}+\right.\\
&\left.2\text{Re}\left[G_{AB}(\mathbf{r})\right]\prod_\mu H_{n_\mu}^{(0)}(\xi_{\mu,A})H_{m_\mu}^{(0)}(\xi_{\mu,A})e^{-\frac{1}{2}(3\xi_{\mu,A}^2+\xi_{\mu,B}^2)}\right],
\label{eq:eq18}
\end{split}
\end{equation}

\begin{equation}
\begin{split}
&O_{\mathbf{n} B,\mathbf{m} B}=b_{x,B}b_{y,B}b_{z,B}\cdot\\
&\int_{\mathbb{R}^3} d\mathbf{r} \left[G_{AA}(\mathbf{r})\prod_\mu H_{n_\mu}^{(0)}(\xi_{\mu,B})H_{m_\mu}^{(0)}(\xi_{\mu,B})e^{-(3\xi_{\mu,A}^2+\xi_{\mu,B}^2)}+\right.\\
&\left.G_{BB}(\mathbf{r})\prod_\mu H_{n_\mu}^{(0)}(\xi_{\mu,B})H_{m_\mu}^{(0)}(\xi_{\mu,B})e^{-2\xi_{\mu,B}^2}+\right.\\
&\left.2\text{Re}\left[G_{AB}(\mathbf{r})\right]\prod_\mu H_{n_\mu}^{(0)}(\xi_{\mu,B})H_{m_\mu}^{(0)}(\xi_{\mu,B})e^{-\frac{1}{2}(\xi_{\mu,A}^2+3\xi_{\mu,B}^2)}\right],
\label{eq:eq19}
\end{split}
\end{equation}

\begin{equation}
\begin{split}
&O_{\mathbf{n} B,\mathbf{m} A}=\sqrt{b_{x,A}b_{y,A}b_{z,A}b_{x,B}b_{y,B}b_{z,B}}\cdot\\
&\int_{\mathbb{R}^3} d\mathbf{r} \left[G_{AA}(\mathbf{r})\prod_\mu H_{n_\mu}^{(0)}(\xi_{\mu,B})H_{m_\mu}^{(0)}(\xi_{\mu,A})e^{-\frac{1}{2}(3\xi_{\mu,A}^2+\xi_{\mu,B}^2)}+\right.\\
&\left.G_{BB}(\mathbf{r})\prod_\mu H_{n_\mu}^{(0)}(\xi_{\mu,B})H_{m_\mu}^{(0)}(\xi_{\mu,A})e^{-\frac{1}{2}(\xi_{\mu,A}^2+3\xi_{\mu,B}^2)}+\right.\\
&\left.2\text{Re}\left[G_{AB}(\mathbf{r})\right]\prod_\mu H_{n_\mu}^{(0)}(\xi_{\mu,B})H_{m_\mu}^{(0)}(\xi_{\mu,A})e^{-(\xi_{\mu,A}^2+\xi_{\mu,B}^2)}\right],
\label{eq:eq20}
\end{split}
\end{equation}

Hence, in Eqs.~(\ref{eq:eq18})-(\ref{eq:eq20}), there appear five different Gaussians, which translates into five different lattices where the quadratures need to be computed. After several algebraic steps, we then define five different scaled coordinates, shown in Table (\ref{Tab:table1}).

\begin{table}[htb]
\begin{center}
\begin{tabular}{||c c||}
 \hline
 $\#A-\#B$ & $\eta^{iji'j'}$ \\ [0.5ex] 
 \hline\hline
 4-0  & $\frac{1}{\sqrt{2}b_{\mu,A}}\eta+r_{\mu 0,A}$ \\ 
 \hline
 3-1  & ${\sqrt\frac{2}{3b_{\mu,A}^2+b_{\mu,B}^2}}\eta+\frac{3b_{\mu,A}^2r_{\mu 0,A}+b_{\mu,B}^2r_{\mu 0,B}}{3b_{\mu,A}^2+b_{\mu,B}^2}$\\
 \hline
 2-2 & $\frac{1}{\sqrt{b_{\mu,A}^2+b_{\mu,B}^2}}\eta+\frac{b_{\mu,A}^2r_{\mu 0,A}+b_{\mu,B}^2r_{\mu 0,B}}{b_{\mu,A}^2+b_{\mu,B}^2}$\\
 \hline
 1-3 & ${\sqrt\frac{2}{b_{\mu,A}^2+3b_{\mu,B}^2}}\eta+\frac{b_{\mu,A}^2r_{\mu 0,A}+3b_{\mu,B}^2r_{\mu 0,B}}{b_{\mu,A}^2+3b_{\mu,B}^2}$\\
 \hline
 0-4 & $\frac{1}{\sqrt{2}b_{\mu,B}}\eta+r_{\mu 0,B}$\\ [1ex] 
 \hline
\end{tabular}
\caption{Different scaling of space coordinates $r_\mu$ for Gaussians appearing in Eqs.~(\ref{eq:eq18}-\ref{eq:eq20}). The first column shows how many times the widths $b_{\mu,A}$ and $b_{\mu,B}$ appear in the quadratures. Factors $\eta^{iji'j'}$ denote variables $\eta$ scaled and shifted depending on the integrands indices $i,j,i',j'$.}
\label{Tab:table1}
\end{center}
\end{table}

To simplify the expressions (and the computational burden), we can rewrite the product of the two Hermite polynomials involved as a finite sum of another Hermite polynomial in the style of~\cite{Dob1997}:

\begin{equation}
\begin{split}
& H_{n_\mu}^{(0)}[b_{\mu,i}(\eta^{iji'j'}_\mu-r_{\mu0,i})]H_{m_\mu}^{(0)}[b_{\mu,j}(\eta^{iji'j'}_\mu-r_{\mu0,j})]=\\
& \sum_{k_\mu=0}^{n_\mu+m_\mu}C_{n_\mu m_\mu k_\mu}^{iji'j'}H^{(0)}_{k_\mu}(\eta_\mu),
\label{eq:eq22}
\end{split}
\end{equation}
where coefficients $C_{n_\mu m_\mu k_\mu}^{iji'j'}$ can be computed numerically via the Gauss-Hermite quadrature again. Taking this fact and the proper algebraic modifications into account, the general TCHO matrix element reads

\begin{equation}
\begin{split}
& O_{\mathbf{n} i,\mathbf{m}j}=\sum_{\substack{i'j' \\ k_x k_y k_z}}\Omega^{iji'j'}C_{n_x m_x k_x}^{iji'j'}C_{n_y m_y k_y}^{iji'j'}C_{n_z m_z k_z}^{iji'j'} O_{k_xk_yk_z}^{iji'j'},
\label{eq:eq23}
\end{split}
\end{equation}
where 
\begin{equation}\label{eq_a7}
    O_{k_xk_yk_z}^{iji'j'}= \int_{\mathbb{R}^3} d\vec{\eta} G_{i'j'}(\vec{\eta}^{iji'j'})\prod_\mu H_{k_\mu}^{(0)}(\eta^{iji'j'}_\mu)e^{-\eta_\mu^2}.
\end{equation}
Coefficients $C_{n_\mu m_\mu k_\mu}^{iji'j'}$ contain hidden factors $\sqrt{b_{\mu,i}b_{\mu,j}}$ and

\begin{equation}
\Omega^{iji'j'}=\prod_{\mu}\sqrt{\frac{2}{b_{\mu,i}^2+b_{\mu,j}^2+b_{\mu,i'}^2+b_{\mu,j'}^2}}e^{-\frac{1}{2}\bar{B}_\mu^{iji'j'}},
\label{eq:eq24}
\end{equation}
where the exponent depends on the HO constants and bases' shifts,
\begin{equation}
\begin{split}
\bar{B}_\mu^{iji'j'}=& b_{\mu,i}^2r_{\mu 0,i}^2+b_{\mu,j}^2r_{\mu 0,j}^2+b_{\mu,i'}^2r_{\mu 0,i'}^2+b_{\mu,j'}^2r_{\mu 0,j'}^2 \\
&-\frac{\left(b_{\mu,i}^2r_{\mu 0,i}+b_{\mu,j}^2r_{\mu 0,j}+b_{\mu,i'}^2r_{\mu 0,i'}+b_{\mu,j'}^2r_{\mu 0,j'}\right)^2}{b_{\mu,i}^2+b_{\mu,j}^2+b_{\mu,i'}^2+b_{\mu,j'}^2}.
\label{eq:eq25}
\end{split}
\end{equation}

The same strategy can be used to evaluate the energy of the functional, where the integrands are products of pairs of densities. Therefore, again, we deal with the products of four Gaussians with different combinations of the HO constants and shifts. However, the lattices to compute the quadratures are the same, and the polynomials required are evaluated again at the same points.
 The only difference is that we need to perform the sum over all four indices instead of computing only one cross-term as before.

\section{Evaluation of densities}
\label{appendixC}

In~\ref{appendixA}, we saw that five different lattices are needed to perform the numerical integration for the matrix elements or energy of the functional. This is because the four indices $i,j,i',j'$ define just five different combinations of shifts and scaling, summarized in Table \ref{Tab:table1}. However, from Eq.~(\ref{eq_a7}), we see that we need to evaluate polynomials $G_{i'j'}$ that define densities in the different lattice points of $\eta^{iji'j'}$.

In the density-dependent (DD) term, the coupling constants are proportional to $\rho_0^{\gamma}(\mathbf{r})$, with $\rho_0(\mathbf{r})=\rho_p(\mathbf{r})+\rho_n(\mathbf{r})$ and $\gamma$ being (usually) a non-integer parameter of the interaction~\cite{Schunck2019}. Therefore, the Gauss-Hermite quadratures are no longer exact because, in general, the integrands are not polynomials but products of polynomials and non-integer powers of polynomials. Since the total densities are relatively smooth functions, quadratures defined for the density-independent terms can still be quite precise. Nevertheless, as discussed below, those terms require special treatment in the implementation of TCHO. 

To determine how many different polynomials $G_{i'j'}$ must be evaluated, we use the following principles:
\begin{itemize}
\item Due to $G_{i'j'}=G_{j'i'}^*$ we only explicitly compute those for $j'\leq i'$. Hence, three different polynomials in five lattices make a total of 15. In the case of the pairing densities, though, this doesn't apply because left and right wave functions are related, respectively, to the lower and upper components of the quasiparticle wave functions~\cite{Dob84}, giving. As a result, 20 different polynomials.
\item The polynomial $G_{i'j'}$ already fixes two of the four indices of the lattice to be used. As a result, if DD terms are not considered, only 3 out of 5 possible lattices must be used. For example, if we take $G_{AA}$, only those lattices with $\#A\geq2$ in Table \ref{Tab:table1} are used (which are 4-0,3-1,2-2). Then, their number can be reduced to 9 polynomials or 12 when including pairing in the calculation.
\end{itemize}

To summarize these results, in Table \ref{Tab:counting}, we show the number of different polynomials needed in the function of the complexity of the calculation.

\begin{table}[h!]\label{Tab:conting}
\centering
\begin{tabular}{||c |c| c||} 
 \hline
  & \makecell{w/o pairing dens. \\ (HF)} & \makecell{ w/ pairing dens. \\ (HFB)} \\ [0.5ex] 
 \hline\hline
 \makecell{no DD terms \\ (SV, SV$_\text{T}$,...)} & 9 & 12 \\ 
 \makecell{with DD terms \\ (SkM*,UDF1,...)} & 15 & 20 \\ [1ex] 
 \hline
\end{tabular}
\caption{Total number of polynomials in the quadrature lattices needed for different methods and interactions.}
\label{Tab:counting}
\end{table}

When using the DD interactions, the so-called rearrangement term is required in the mean field, whose structure is different from the rest of the terms~\cite{Dob1997}:
\begin{equation}
    U^{\text{rear}}(\mathbf{r})\propto \gamma \rho_{0}^{\gamma-1}(\mathbf{r})\rho^2(\mathbf{r}),
    \label{eq:eq34}
\end{equation}
where to simplify the presentation, we omitted the part related to the spin density. Then, the matrix elements of the rearrangement term read
\begin{equation}
(U^{\text{rear}}_{ij})_{\mathbf{nm}}\propto \gamma \int_{\mathbb{R}^3} d\mathbf{r} \phi_{\mathbf{n},i}^*(\mathbf{r}) \rho_{0}^{\gamma-1}(\mathbf{r})\rho^2\phi_{\mathbf{m},j}(\mathbf{r}).
\label{eq:eq35}
\end{equation}

One last remark related to the numerical evaluation of this integral is in order. In the calculations performed in the TCHO basis, we can encounter situations where the true density (not its polynomial part) is extremely small. This is the case, for instance, when the two centers are separated to describe the fission fragments far away from one another. In that case, the rearrangement term can lead to numerical issues when the power $\gamma$ of the interaction is smaller than one due to the division by extremely small values. To avoid this behavior, the rearrangement term must be set to zero at small densities. 

\section{TCHO matrix elements of the Coulomb interaction}
\label{appendixD}

To compute the integrals appearing in (\ref{eq:eq65}) we need to transform the expression in such a way that $r_{\mu 1}$ and $r_{\mu 2}$ are separable. Let's treat only the exponent, recalling:

\begin{equation}
\begin{split}
&E(r_{\mu 1},r_{\mu 2})=\\
& \frac{1}{2}\left[(\xi_{1\mu,i}^2+\xi_{1\mu,i'}^2)+(\xi_{2\mu,j}^2+\xi_{2\mu,j'}^2)+2a_\gamma(r_{\mu 1}-r_{\mu 2})^2\right],
\label{eq:eq66}
\end{split}
\end{equation}
so that, the total Gaussian function in $(\ref{eq:eq65})$ is simply $e^{-E(r_{\mu 1},r_{\mu 2})}$. Expanding all the terms involved in (\ref{eq:eq66}), we have the following quadratic form 

\begin{equation}\label{eq:eq67}
    E(r_{\mu 1},r_{\mu 2})=ar_{\mu 1}^2+br_{\mu 2}^2+cr_{\mu 1}r_{\mu 2}+dr_{\mu 1}+er_{\mu 2}+f,
\end{equation}
where
\begin{subequations}\label{eq:eq68}
   \begin{equation}
      a=\frac{1}{2}(b_{\mu,i}^2+b_{\mu,i'}^2+2a_\gamma),
\end{equation}
   \begin{equation}
      b=\frac{1}{2}(b_{\mu,j}^2+b_{\mu,j'}^2+2a_\gamma),
   \end{equation}
   \begin{equation}
      c=-2a_\gamma,
   \end{equation}
   \begin{equation}
      d=-(b_{\mu,i}^2r_{\mu 0,i}+b_{\mu,i'}^2r_{\mu 0,i'}),
   \end{equation}
   \begin{equation}
      e=-(b_{\mu,j}^2r_{\mu 0,j}+b_{\mu,j'}^2r_{\mu 0,j'}),
   \end{equation}
   \begin{equation}
      f=\frac{1}{2}(b_{\mu,i}^2r_{\mu 0,i}^2+b_{\mu,i'}^2r_{\mu 0,i'}^2+b_{\mu,j}^2r_{\mu 0,j}^2+b_{\mu,j'}^2r_{\mu 0,j'}^2),
   \end{equation}
\end{subequations}
which can be transformed into a new one of the type
\begin{equation}
E'\left(\eta_1,\eta_2\right)=A\eta_1^{2}+B\eta_2^{2}+C.
\label{eq:eq69}
\end{equation}

For that purpose, let's write (\ref{eq:eq67}) in matrix notation:
\begin{equation}\label{eq:eq70}
    E(r_{\mu 1},r_{\mu 2})=\mathbf{x}^T A\mathbf{x} +B \mathbf{x} + f,
\end{equation}
where 
\begin{subequations}\label{eq:eq71}
\begin{equation}
\mathbf{x}=
\begin{pmatrix}
r_{\mu 1} \\
r_{\mu 2}
\end{pmatrix},
\end{equation}
\begin{equation}
A=
\begin{pmatrix}
a & c/2\\
c/2 & b
\end{pmatrix}, \ \ 
B=
\begin{pmatrix}
d & e 
\end{pmatrix}
\end{equation}
\end{subequations}
then, we can apply a linear transformation to eliminate the cross-term in the coordinates (proportional to $r_{\mu 1}r_{\mu 2}$). If we insert the next transformation  

\begin{equation}\label{eq:eq72}
    \mathbf{x}=P\mathbf{x'} \longrightarrow
    \begin{pmatrix}
r_{\mu 1} \\
r_{\mu 2}
\end{pmatrix}=
\begin{pmatrix}
P_{11} & P_{12} \\
P_{21} & P_{22}
\end{pmatrix}
    \begin{pmatrix}
r_{\mu 1}' \\
r_{\mu 2}'
\end{pmatrix}
\end{equation}
into (\ref{eq:eq67}) we have,
\begin{equation}\label{eq:eq73}
    E'(r_{\mu 1}',r_{\mu 2}')=\mathbf{x'}^T A' \mathbf{x'} +B'\mathbf{x'} + f,
\end{equation}
where $A'$ is diagonal. In other words, $P$ is the unitary matrix that diagonalizes $A$. Then, in terms of its eigenvalues,

\begin{subequations}
\begin{equation}\label{eq:eq74}
    A'= P^TAP =  \begin{pmatrix}
\pi_1 & 0 \\
0 & \pi_2
\end{pmatrix},
\end{equation}
\begin{equation}\label{eq:eq75}
B'= BP =  \begin{pmatrix}
d' & e'
\end{pmatrix},
\end{equation}
\end{subequations}
where
\begin{subequations}\label{eq:eq76}
\begin{equation}
d'=P_{11}d+P_{21}e,
\end{equation} 
\begin{equation}
e'=P_{12}d+P_{22}e.
\end{equation} 
\end{subequations}

Then, expression (\ref{eq:eq73}) reads as
\begin{equation}\label{eq:eq77}
E'(r_{\mu 1}',r_{\mu 2}')=\pi_1r_{\mu 1}^{'2}+\pi_2r_{\mu 2}^{'2}+d'r_{\mu 1}'+e'r_{\mu 2}'+f.
\end{equation}

Going back to equation (\ref{eq:eq65}) and taking into account that $P$ is orthogonal --as it diagonalizes the symmetric matrix $A$-- we can change the variables of integration as follows

\begin{equation}
\begin{split}
    & v^\gamma_{n_{i}m_{j}n'_{i'}m'_{j'}}=\\
    &e^2 A_\gamma\sqrt{b_{\mu,i}b_{\mu,j}b_{\mu,i'}b_{\mu,j'}} \iint dr_{\mu 1}' dr_{\mu 2}'\cdot\\
    & H_{n_\mu}^{(0)}(\xi_{1\mu,i})H_{m_\mu}^{(0)}(\xi_{2\mu,j})H_{n'_\mu}^{(0)}(\xi_{1\mu,i'})H_{m'_\mu}^{(0)}(\xi_{2\mu,j'})e^{-E'(r_{\mu 1}',r_{\mu 2}')},
\end{split}
\label{eq:eq78}
\end{equation}
If one finally completes the squares and makes the right changes of variables, the matrix elements can be computed numerically with the double quadrature as
\begin{equation}
\begin{split}
   & v^\gamma_{n_{i}m_{j}n'_{i'}m'_{j'}}= e^2 A_\gamma T_{ii'jj'}  \sum_{\alpha\beta} w_{\alpha}w_{\beta} \cdot \\ & H_{n_\mu}^{(0)}(\eta_{1,\alpha},\eta_{2,\beta})H_{m_\mu}^{(0)}(\eta_{1,\alpha},\eta_{2,\beta})\\
   & H_{n'_\mu}^{(0)}(\eta_{1,\alpha},\eta_{2,\beta})H_{m'_\mu}^{(0)}(\eta_{1,\alpha},\eta_{2,\beta}),
\end{split}
\label{eq:eq85}
\end{equation}
where
\begin{subequations}
\begin{equation}
T_{ii'jj'}=\sqrt{\frac{b_{\mu,i}b_{\mu,j}b_{\mu,i'}b_{\mu,j'}}{\pi_1\pi_2}}e^{-f'}, 
\end{equation}
\begin{equation}
f'=f-\frac{d^{'2}}{4\pi_1}-\frac{e^{'2}}{4\pi_2},
\label{eq:eq81}
\end{equation}
\end{subequations}
and the new lattice is related to the rotated coordinates as

\begin{subequations}\label{eq:eq82}
    \begin{equation}
        \eta_1=\sqrt{\pi_1}r_{\mu 1}'+\frac{d'}{2\sqrt{\pi_1}},
    \end{equation}
    \begin{equation}
        \eta_2=\sqrt{\pi_2}r_{\mu 2}'+\frac{e'}{2\sqrt{\pi_2}}.
    \end{equation}
\end{subequations}

\bibliographystyle{spphys}       
\bibliography{ref}   

\end{document}